\documentclass[final,3p,times]{elsarticle}

\usepackage{multirow}
\usepackage{graphicx}
\usepackage[table,xcdraw]{xcolor}
\usepackage[linesnumbered,ruled,vlined,boxed]{algorithm2e}
\usepackage{amsmath}
\usepackage{amssymb}
\usepackage{verbatim}
\usepackage{float}
\usepackage{subcaption} 
\usepackage{lineno}
\usepackage{booktabs}

\biboptions{comma,square,sort&compress}

\journal{Speech Communication}

\begin{document}

\begin{frontmatter}


\title{Acoustic Characterization and Machine Prediction of Perceived Masculinity and Femininity in Adults}


\author[label1]{Fuling Chen}
\ead{fuling.chen@uwa.edu.au}
\author[label1]{Roberto Togneri}
\ead{roberto.togneri@uwa.edu.au}
\author[label2]{Murray Maybery}
\ead{murray.maybery@uwa.edu.au}
\author[label2,label3]{Diana Tan}
\ead{diana.tan@uwa.edu.au}
\address[label1]{Dept. of Electrical, Electronic and Computer Engineering, University of Western Australia}
\address[label2]{School of Psychological Science, University of Western Australia}
\address[label3]{Telethon Kids Institute, Perth, Australia}

\begin{abstract}
Previous research has found that voices can provide reliable information to be used for gender classification with a high level of accuracy. In social psychology, perceived masculinity and femininity (masculinity and femininity rated by humans) has often been considered an important feature when investigating the influence of vocal features on social behaviours. While previous studies have characterised the acoustic features that contributed to perceivers’ judgements of speakers’ masculinity or femininity, there is limited research on developing a machine masculinity/femininity scoring model and characterizing the independent acoustic factors that contribute to perceivers’ masculinity and femininity judgements. In this work, we first propose a machine scoring model of perceived masculinity/femininity based on the Extreme Random Forest and then characterize the independent and meaningful acoustic factors that contribute to perceivers’ judgements by using a correlation matrix based hierarchical clustering method. Our results show that the machine ratings of masculinity and femininity strongly correlated with the human ratings of masculinity and femininity when we used an optimal speech duration of 7 seconds, with a correlation coefficient of up to .63 for females and .77 for males. Nine independent clusters of acoustic measures were generated from our modelling of femininity judgements for female voices and eight clusters were found for masculinity judgements for male voices. The results revealed that, for both genders, the F0 mean is the most important acoustic measure affecting the judgement of acoustic-related masculinity and femininity. The F3 mean, F4 mean and VTL estimators were found to be highly inter-correlated and appeared in the same cluster, forming the second most significant factor in influencing the assessment of acoustic-related masculinity and femininity. Next, F1 mean, F2 mean and F0 standard deviation are independent factors that share similar importance. The voice perturbation measures, including HNR, jitter and shimmer, are of lesser importance in influencing masculinity/femininity judgements.
\end{abstract}


\end{frontmatter}


\section{Introduction}

The human voice varies on a range of characteristics including acoustic cues (e.g., pitch), linguistics (e.g., vocabulary) and paralinguistics (e.g., vocal expressions). These vocal characteristics have been implicated in listeners’ judgements of vocal masculinity or femininity \cite{cartei2014makes,feinberg2006menstrual,feinberg2005manipulations,pisanski2011prioritization,ko2006voice,hardy2020acoustic,gelfer2000comparison,munson2007acoustic,king2012voice,biemans2000gender,owen2010role,nolan2019role}, which, in turn, have been shown to influence listeners’ evaluations of the speakers’ sex \cite{ko2006voice} and personality 
\cite{ko2009stereotyping}. For instance, more masculine voices are rated more highly for social dominance for both male and female speakers \cite{jones2010domain,puts2006dominance}. Gender identity refers to the extent to which individuals perceive themselves to be masculine or feminine based on the social expectation of being a man or woman in their society \cite{stets2000femininity} \footnote{Note that gender identity is distinct from gender roles and gender stereotypes, where gender roles are shared expectations of behavior given one's gender \cite{eagly2013sex}, and gender stereotypes are shared views of personality traits such as instrumentality in men and expressiveness in women \cite{helmreich1978achievement}}. Similarly, perceived masculinity or femininity refers to the degree to which persons are regarded by the society members as masculine or feminine. Gender identity is not restricted to being either a man or a woman. Other common identities also exist, such as LGBTIQA+ (lesbian, gay, bisexual, transgender, intersex, queer/questioning, asexual and other identities). Perceived masculinity/femininity could vary in specific ways among the various gender identities. As a first step in this investigation, we chose to focus on the binary biological sexes and examine the sources of variation in human-rated masculinity/femininity within each class, focusing particularly on the contributing acoustic factors for each sex. One reason for taking this approach is that the binary biological sexes are more definitive than the various gender identities, and the underlying differences between different forms of gender identities are often nuanced and highly variable across individuals. A second reason is that sex-linked biological factors influence development of the voice organ which in turn influences the acoustics of the voice and perceived masculinity/femininity (see below and \cite{zamponi2021effect}).  Understanding of this causal pathway can be enhanced by investigating the acoustic features that influence the perceived masculinity and femininity of male and female voices. In pursuing this investigation, we recognize that perceived masculinity and femininity are influenced by other features of verbal communication such as linguistics (e.g., vocabulary \cite{walker2022s}) and paralinguistics (e.g., facial expression, gesture \cite{kawamura2008smiling}), and also by body morphology (e.g., posture, facial structure \cite{lefevre2014perceiving}) and appearance (e.g., hairstyle and clothing \cite{bereczkei2006hair}). 

In general, differences in perceived masculinity and femininity are associated with differences between males and females in the development of secondary sex characteristics \cite{andersson2019sexual}, such as the length of the vocal tract \cite{fitch1999morphology}. These sex differences are influenced by biological factors (e.g., higher levels of masculinizing hormones like testosterone). Furthermore, variations in secondary sex characteristics have been found to correlate with health status, physical strength, mating success and the presence of psychological disorders \cite{fink2007male,honekopp2007physical,samson2000relationships}. Several studies have shown that perceived masculinity and femininity play important roles in human social preferences such as attractiveness \cite{cartei2014makes,feinberg2006menstrual,little2011human,feinberg2008correlated,feinberg2005manipulations,pisanski2011prioritization}. There are two sources of information that could influence one's judgements of masculinity/femininity, which are the structure of the face and the characteristics of the voice. While the influence of facial structure on perceived masculinity/femininity has been investigated in some studies \cite{feinberg2006menstrual,little2011human,feinberg2008correlated} there has been less research investigating the relationships between a wide range of acoustic features of voice and the perceived masculinity/femininity. The present study aimed to comprehensively investigate those relationships.

Some existing studies have investigated the relationships between the perceived masculinity/femininity of voices and various acoustic measures, as summarized in Table \ref{tab:intro_relatedworks_summary}. Cartei et al. \cite{cartei2014makes} examined relationships between acoustic measures (e.g. fundamental frequency (F0) and resonance ($\Delta$F)) and masculinity judgements of male voices. Male speakers with lower F0 and $\Delta$F were rated as more masculine, with F0 showing as a more salient cue for perceived masculinity than $\Delta$F. Similarly, Ko et al. \cite{ko2006voice} demonstrated that F0 was highly correlated with femininity judgements of male (r = -.59) and female (r = -.69) voices. Furthermore, the study showed that $\Delta$F and F0 variance (F0SD) had weaker correlations with femininity for both males (r($\Delta$F,femininity) = -.14, r(F0SD,femininity) = .46) and females (r($\Delta$F,femininity) = .28, r(F0SD,femininity) = .26), compared to the correlation between F0 and femininity; F0SD was related to vocal femininity only in males and not in females. Other acoustic features associated with perceived masculinity and femininity have also been reported in the literature. For instance, perceived masculinity/femininity was found to be related to F0 and formant frequencies (Fn) for the binary sexes \cite{pisanski2011prioritization,hardy2020acoustic} and also for transgender groups \cite{hardy2020acoustic,gelfer2000comparison,king2012voice,owen2010role}. Specifically, higher F0 and higher Fn were associated with perceiving voices as more feminine and less masculine. Fn variables have also been shown to correlate with perceived masculinity/femininity ratings for both sexes, with the second formant frequency (F2) correlating more strongly with masculinity/femininity ratings than either the first (F1) or the third (F3) formant frequency \cite{gelfer2000comparison}. Of a range of acoustic measures examined in \cite{munson2007acoustic}, F0 and F2 were found to account for 49.60\% and 19.60\% of the variance in masculinity ratings for males respectively, and for 40.60\% and 24.10\% of the variance in femininity ratings for females respectively. It was found in the systematic review of \cite{nolan2019role} that, through phonosurgery and voice therapy, F0 could be increased significantly for male-to-female transgender individuals, which resulted in an increase in perceived vocal femininity. Feinberg et al. \cite{feinberg2006menstrual,feinberg2005manipulations} showed that F0 and apparent vocal tract length (VTL) both influenced the masculinity ratings for male speakers. The voice perturbation measures, including jitter, shimmer and Harmonic-to-Noise Ratio (HNR), have also been investigated in their associations with perceived masculinity and femininity. In the studies of \cite{hardy2020acoustic,king2012voice,owen2010role}, the authors found that voice perturbation measures did not correlate with perceived masculinity and femininity, whereas Biemans \cite{biemans2000gender} found that the correlation between HNR and femininity ratings for females (r = .29) was stronger than that of HNR and masculinity ratings for males (r = -.08). In summary, the existing literature provides strong evidence for F0 as a critical measure in influencing perceived masculinity/femininity, with other acoustic measures such as F0 variance, Fn, $\Delta$F, VTL, HNR, jitter and shimmer as potential valid cues to human listeners in their judgement of perceived masculinity and femininity expressed in the speakers. 

\begin{table}[h!]
	\caption{Summary of research on acoustic measures in perceived masculinity and femininity}
	\label{tab:intro_relatedworks_summary}
	\centering
	\resizebox{\columnwidth}{!}{
		\begin{tabular}{p{0.08\linewidth}p{0.25\linewidth}p{0.15\linewidth}p{0.15\linewidth}p{0.15\linewidth}p{0.1\linewidth}p{0.2\linewidth}p{0.4\linewidth}}
			\toprule[0.4ex]
			\textbf{Reference}&\textbf{Acoustic Measures}&\textbf{Subjects}&\textbf{Raters}&\textbf{Judgements}&\textbf{Method+}&\textbf{Stimuli Type}&\textbf{Findings}\\\midrule[0.4ex]
			\cite{cartei2014makes}&F0, $\Delta$F&37M*&20F*&Mas$\ddagger$ for M&1
			&Isolated word; sentence; connected speech&F0 mediated correlation between perceived masculinity and testosterone levels.\\\midrule
			\cite{ko2006voice}&F0, $\Delta$F, SD of $\Delta$F&47M, 47 F&24M, 30F&Fem$\ddagger$ for M, F&1
			&Passage reading&F0 was highly and positively correlated with femininity for males and females.\\\midrule
			\cite{feinberg2006menstrual,feinberg2005manipulations}&F0, VTL&4M, 4F\cite{feinberg2006menstrual}; 10M\cite{feinberg2005manipulations}&26F\cite{feinberg2006menstrual}; 89F\cite{feinberg2005manipulations}&Fem for M, F&2&Vowels&F0 and VTL were independent and correlated with perceived masculinity.\\\midrule
			\cite{little2011human}&F0&10M, 10F&24M, 25F&Mas for M, fem for F&2&Not mentioned&Men with lower F0 were rated more masculine; women with higher F0 were rated more feminine.\\\midrule
			\cite{pisanski2011prioritization}&F0 and Fn (F1, F2, F3, F4)&57M, 57F&30M, 31F&Mas for M, fem for F&2&Single-syllable bVt words&F0 and Fn were independent and correlated with perceived masculinity.\\\midrule
			\cite{gelfer2000comparison}&F0, F0 range, intonation for two sentences samples; F1, F2, F3 for vowels&15T*, 3M, 6F&20&F-M$\ddagger$ for T, M and F&1&Two sentences; vowels each for 5 seconds&F2 was more statistically significant in perceived masculinity/femininity ratings than either F1 or F3.\\\midrule
			\cite{munson2007acoustic}$\diamond$&F0, F1, F2, F0 range, VTL, /s/ center of gravity, /s/ skew-ness, H2-H1 amplitude for /æ/&11M, 11F&40&Mas for M, Fem for F&3&Words&F0 and F2 were significant predictors of perceived masculinity in males and perceived femininity in females.\\\midrule
			\cite{hardy2020acoustic}&F0, average formant frequency, shimmer, HNR&22T*, 10F, 10M&10M, 10F&M-F$\ddagger$ for T, F and M&1&5 seconds, vowels&Higher F0 and average formant frequency were more strongly associated with feminine ratings than masculinity ratings\\\midrule
			\cite{owen2010role,king2012voice}&F0, HNR, jitter, shimmer \cite{owen2010role}; F0, Fn, jitter, shimmer \cite{king2012voice}&20T, 5F, 5M \cite{owen2010role}; 21T, 9F \cite{king2012voice}&12M, 13F \cite{owen2010role}; 15F, 5M \cite{king2012voice}&Fem for T, F and M\cite{owen2010role}; Fem-Mas for T and F\cite{king2012voice}&1 \cite{owen2010role}; 4 \cite{king2012voice}&Passage reading, 20 - 25 seconds&F0 was strongly correlated with speaker's self-rated and listener-rated femininity, but HNR, jitter and shimmer did not correlate with these ratings.\\\midrule
			\cite{biemans2000gender}&F0, HNR, jitter, shimmer&57M, 57F&Self-assessment&Mas for M, Fem for F&1&45 seconds of speech reduced from 30-minute interactions&HNR was found to be correlated more strongly with femininity ratings for females than with masculinity ratings in males.\\
			\bottomrule[0.4ex]
			\multicolumn{8}{p{1.6\linewidth}}{* M - self-reported males; F - self-reported females; T - transgender women who were identify as transsexual or transgender and to be living in their affirmed gender role (i.e., woman) the majority of the time (ie, at least 80\% of the time) for at least 6 months.}\\
			\multicolumn{8}{p{1.6\linewidth}}{$\diamond$ This study also investigated gay men and lesbian, but the speakers were rated in a different scale.}\\
			\multicolumn{8}{p{1.6\linewidth}}{$\ddagger$ Mas - masculinity ratings on a scale from not at all masculine to very masculine; Fem - femininity ratings on a scale from not at all feminine to very feminine; Fem-Mas - a rating scale from very feminine to very masculine; Mas-Fem - a rating scale of very masculine to very feminine}\\
			\multicolumn{8}{p{1.6\linewidth}}{+ 1. Correlation of acoustic measures and perceived masculinity/femininity ratings; 2. Manipulation of acoustic characteristics; 3. Characterization using regression models; 4. Statistical analysis using t-test.}\\
		\end{tabular}
	}
\end{table}

However, there are two key limitations in the previous studies. Firstly, as listed in the column ``Acoustic Measures" in Table \ref{tab:intro_relatedworks_summary}, each previous study has only investigated a small set of acoustic measures, which limits a thorough understanding of how the broader range of acoustic factors contribute to judgements of perceived masculinity and femininity. Secondly, in some of the previous studies, such as \cite{pisanski2011prioritization,hardy2020acoustic,owen2010role,king2012voice,biemans2000gender}, the inter-correlations among the assessed acoustic measures were not considered. Though it would not affect the analysis of the association between each acoustic measure and the perceived judgements, high levels of inter-correlation between acoustic measures may reflect dependencies among the acoustic measures, thus limiting our capability to interpret the importance of several acoustic measures.

To address these limitations, we conducted a study investigating the utility of a comprehensive set of acoustic measures in modeling masculinity ratings for males and femininity ratings for females, using a methodology that addressed the issue of inter-correlations among the acoustic variables. In this work we developed a machine learning model that may provide an invaluable analysis and diagnostic tool for psychological researchers and clinicians in their assessment of masculinity and femininity using acoustic correlates. We deployed machine learning models in the present study for the following reasons: (1) none of the above-mentioned studies suggest how to design a machine system that could be used to predict an unknown speaker's perceived masculinity or femininity which would be an invaluable aid; and (2) traditional analysis methods cannot provide solutions to automatically address the complex inter-correlations among multiple acoustic measures. 

The aim of this study is to build a machine rating model of masculinity and femininity based on a comprehensive set of acoustic measures which can be used to characterize and predict listeners' perceived voice masculinity/femininity. In investigating the viability of a machine rating model of masculinity and femininity, our research used a database of voices that is substantially larger than the databases used in previous research (see Table \ref{tab:intro_relatedworks_summary} where the largest existing database consisted of 114 speakers \cite{biemans2000gender}). The present study used a new dataset of speech segments from 225 adult speakers which were rated for the speakers' perceived masculinity and femininity by 25-30 listeners. This large dataset enabled rigorous testing for our model development. 

We now review the literature on the latest contribution of machine learning in the prediction and characterization of perceived masculinity and femininity, state the current limitations in our understanding of perceived masculinity and femininity prediction and characterization, and motivate our proposed solutions.

\subsection{Machine Learning in Prediction of Perceived Masculinity and Femininity}
\label{txt:Intro_ml_prediction}
In the past two decades, the development of machine learning models has advanced the field of speech science, offering the potential to overcome the limitation of relying on human ratings for prediction and analysis of speaker traits and pathologies. Reliance on human participants or experts is time consuming and expensive. Machine learning has demonstrated powerful analytical and predictive potential across all fields in the medical and psychological speech sciences. Machine learning models are capable of addressing several challenges including: (1) the gender classification problem - automatically predicting the gender identity of adult speakers with binary sexes (males and females)\footnote{The term ``gender classification" refers to the binary classification of males and females. Research on this classification has typically investigated individuals who self-identify as male or female.}, (2) applying machine learning models to rate speakers’ masculinity/femininity, and (3) characterizing cues that affect the determination of perceived masculinity/femininity.

Several studies have addressed the first issue of gender classification based on human speech acoustic parameters for adults using speech signals can achieve an accuracy of approximately 95\%, by applying machine learning classifiers designed for voice-based gender classification, including Random Forest (RF), Linear Discriminant Analysis (LDA), and K-Nearest Neighbour (KNN) \cite{sedaghi2009comparative,raahul2017voice} methods. RF-based models have been found to be a powerful tool for both classification and regression purposes \cite{svetnik2003random}, with multiple applications, such as emotion recognition \cite{li2015simulated} as well as gender classification \cite{ramadhan2017parameter}. Harb et al. \cite{harb2005voice} proposed a set of neural networks using acoustic and pitch-related features for gender classification and achieved 98.5\% accuracy. 

Considering the length of speech to use in either gender classification or in finding relationships to masculinity/femininity ratings, long-term speech was found to be more suitable than short-term speech. Harb et al. \cite{harb2005voice} found that classification was more accurate using segments of 5 seconds compared to 1- and 3-second segments (98.5\% vs 90\% and 93\%). Similarly, Cartei et al. \cite{cartei2014makes} demonstrated that correlations between perceived masculinity for males and both F0 and $\Delta$F were of larger magnitude when the voice samples were connected speech of multiple sentences rather than word-level speech or single-sentence speech. 

To the best of our knowledge, limited research has addressed the second and third challenges of developing machine learning models to rate masculinity and femininity and characterizing the salient cues affecting perceived masculinity/femininity. In the computer vision field, Gilani et al. \cite{gilani2014geometric} proposed a gender classification model based on the LDA algorithm using facial cues. Facial masculinity/femininity scores, generated by the algorithm, significantly correlated with perceived masculinity/femininity scores for males (r = .79) and for females (r = .90). Similar to Gilani's study, in our previous work \cite{chen2020objective}, we designed a machine learning model to rate speakers' acoustic-related masculinity and femininity based on gender classification. We did this by deriving the masculinity/femininity scores for individuals representing where they were positioned in the classification space between extreme ``maleness" (the male model) and extreme ``femaleness" (the female model). The results of our study demonstrated a close correspondence between the machine ratings and human ratings of masculinity for males (r = .67), and femininity for females (r = .51). Besides the LDA algorithm, RF-based regression algorithms have also been widely used to predict human ratings on recommendation of movies \cite{ajesh2016random} and on word prominence judgments \cite{baumann2018makes}. Regarding the characterization of the salient features in prediction, RF-based models have been popularly used in analysing important features in voice-based emotion recognition \cite{rong2009acoustic,cao2017speaker}, detection of Parkinson’s Disease \cite{vaiciukynas2016fusing} and sleep stages classification \cite{xiao2013sleep}.

In the present study, rather than training the model for gender classification as in the previous studies \cite{gilani2014geometric,chen2020objective}, we directly train the model on the human perceptual ratings. We propose a novel masculinity/femininity rating model based on the Extreme Random Forest (ERF) algorithm for predicting human perception of the degree of masculinity of males and the degree of femininity of females, given a comprehensive set of acoustic measures derived from recordings of passage reading and listeners' masculinity/femininity ratings of those recordings. 

\subsection{Machine Learning in Characterization of Perceived Masculinity and Femininity}
\label{txt:Intro_ml_chara}
For the purpose of characterisation, it has been noted that severe multicollinearity increases the difficulty of interpreting regression results \cite{farrar1967multicollinearity,paul2006multicollinearity}. Multicollinearity occurs when two or more predictors are correlated. This correlation is a problem because ideally predictors should be independent. High levels of multicollinearity can prevent interpreting the model results properly to determine the most influential predictors.

Several studies have been conducted to investigate the inter-correlations among various acoustic measures. Cartei et al. \cite{cartei2014makes}, consistent with other studies \cite{fitch1999morphology,fant1970acoustic,fitch2000evolution}, showed that F0 and $\Delta$F are largely independent of each other as they were affected by different constraints on the speech production system. Apart from $\Delta$F, formant dispersion, as another estimator of VTL was shown to be independent of F0 \cite{feinberg2006menstrual,feinberg2005manipulations}. Regarding the relationship between F0 and Fn, the results of Pisanski et al. \cite{pisanski2011prioritization} were consistent with other research \cite{fant1970acoustic,muller1848physiology,macdonald2011probing} in that F0 and Fn were largely independent, both within and across utterances by the same speaker and in the average values of these measures in different speakers. Other studies showed strong inter-correlations existed in some other acoustic measures. For instance, higher formant frequencies, such as F3 and F4, were shown to be strongly correlated with VTL estimators \cite{monahan2010auditory,claes1998novel,tecumseh2001descended}. Severe multicollinearity may also occur with the sources of periodicity perturbations in voice speech signals, including HNR, jitter and shimmer. Some studies showed HNR depended on both jitter and shimmer \cite{krom1993cepstrum,muta1988pitch,qi1997temporal}. However, the inter-correlation may vary between males and females, such that jitter can be independent of other measures in males, while moderately correlated with HNR in females \cite{teixeira2014jitter,lovato2016multi}. The above research has provided evidence of the existence of multicollinearity between selected acoustic measures. Most research works have focused on small sets of acoustic measures and this prevents a more comprehensive and thorough examination of the inter-correlations of all the acoustic measures that contribute to perceived masculinity of males voices and femininity of female voices. However it would be expected that multicollinearity would be worse and more complex to correct if a wide range of acoustic measures were to be considered.

Given the existence of multicollinearity, several approaches have been proposed to reduce the degree of multicollinearity, such as adopting ridge regression and principal components regression \cite{paul2006multicollinearity}. However, a major limitation of ridge regression is that the choice of the biasing constant $k$ is a subjective one and the exact distributional properties are not known \cite{neter1983simultaneous,myers1990classical}. As the degree of multicollinearity is likely to vary among the multiple acoustic measures, the choice of $k$ may not be ideal for all combinations of the acoustic measures. It may cause over-reduction for weakly correlated acoustic measures but under-reduction for highly correlated measures. Principal component regression addresses multicollinearity by using less than the full set of principal components to explain the variation in the response variable. However, the principal components lose the original natural meanings of the variables, so this method is not ideal for characterization purposes. Therefore, in research by Ketchen \cite{ketchen1996application}, clustering methods were designed to address the multicollinearity problem while retaining the natural meanings of variables. In the study of hierarchical cluster analysis \cite{bridges1966hierarchical}, a recommendation was made to cluster the variables with the highest average inter-correlations in the correlation matrix. This correlation matrix based hierarchical clustering method was shown to have higher sensitivity than a method using independent component analysis (ICA), for identifying correlation structures with relatively weak connections, and its outcomes are easy to interpret as the strength of functional connectivity \cite{liu2012correlation}. The present study proposes a method using a novel computational clustering framework to address the multicollinearity issue, which retains the physical meanings of each independent group of acoustic measures. Applying this method prior to using the Extreme Random Forest (ERF) model for prediction, we also propose a new approach to characterize the salient independent groups of acoustic measures that dominate the prediction of perceived masculinity in males and femininity in females.

The remainder of this paper is organized as follows: Section 2 describes the methodology including the datasets and acoustic measures used, the pre-processing of the data and labels, the proposed machine-learning based masculinity/femininity scoring model, the solution of the multicollinearity problem, the acoustic characterization and our evaluation methods. Section 3 presents the results of the modelling and discussion of the outcomes. Finally, Section 4 draws some conclusions based on the study.

\section{Methodology}
\begin{figure}[h!]
	\centering
	\includegraphics[width=0.8\textwidth]{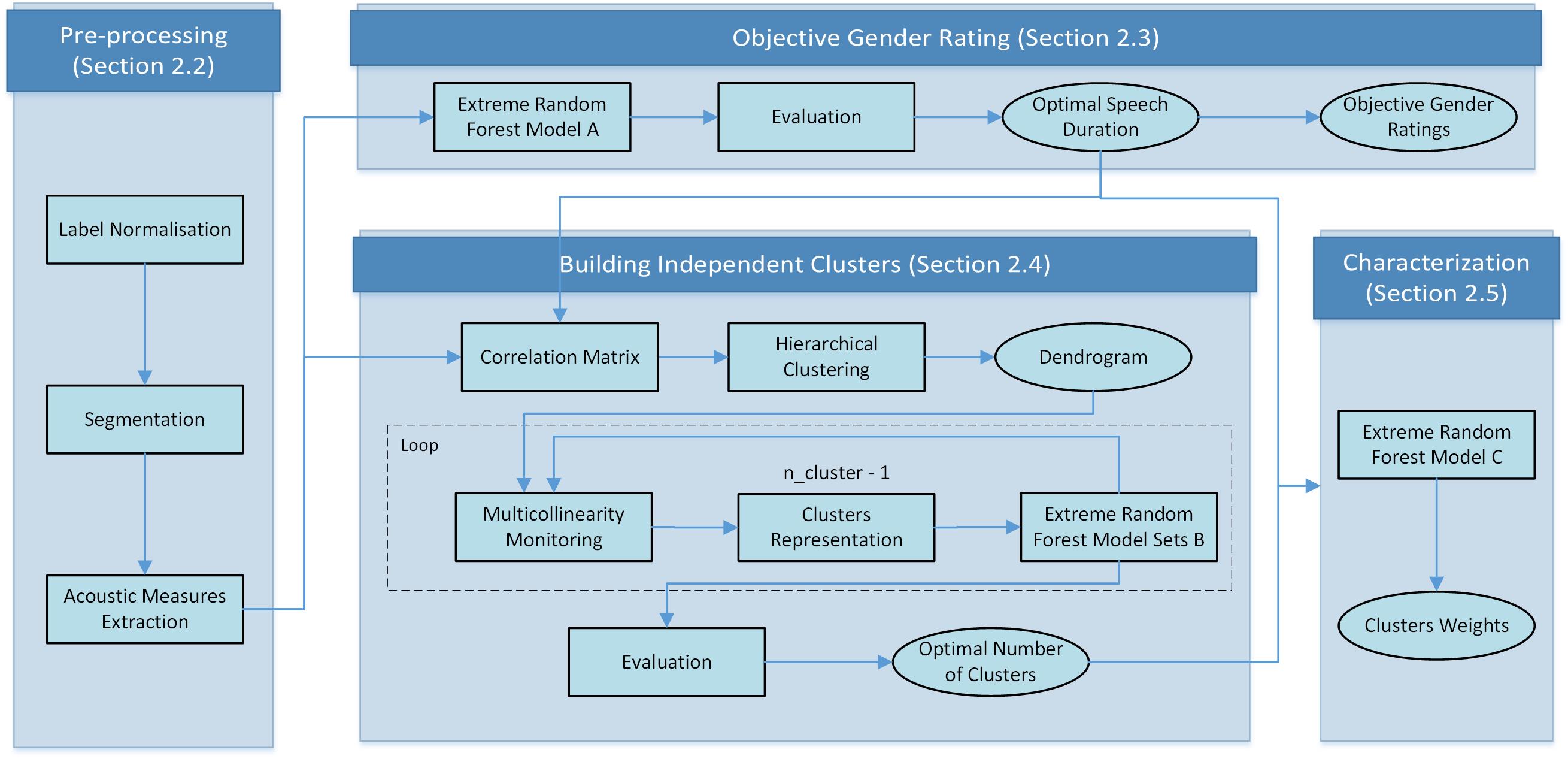}
	\caption{Block diagram of the proposed system}
	\label{fig:method_blockdiagram}
\end{figure}

The proposed system (see Figure \ref{fig:method_blockdiagram}) achieves the following three goals: (1) generation of machine ratings of masculinity/femininity by training on a set of acoustic measures with known perceived masculinity/femininity ratings for both classes (males and females); (2) building independent clusters of acoustic measures with their clearly interpretable meanings for males and females to eliminate multicollinearity; and (3) characterization of the salient clusters of acoustic measures that are associated with perceived masculinity/femininity ratings for both classes. In the remainder of this section we will describe the methods behind each of the goals based on the application of ERF to carry out the masculinity/femininity rating, a novel hierarchical clustering of feature correlations to build the meaningful clusters, and the characterization of the independent clusters.

\subsection{Preparations}
\label{txt:method_preparations}
\subsubsection{Datasets}
\label{txt:method_preparations_datasets}
The datasets were obtained from the School of Psychological Science at the University of Western Australia. Voice recordings were collected for the purpose of investigating the association between perceived masculinity/femininity ratings and autistic traits \cite{tan2015perceived}. This database was chosen because it contains more speakers with available perceived masculinity/femininity ratings than any other public database.

The database (see description in Table \ref{tab:method_preparations_datasets}) is composed of two adult Caucasian cohorts totaling 225 participants (96 self-reporting as male and 129 self-reporting as female with reference to their sex as assigned at birth) who were undergraduate students fluent in English. Tested individually in a soundproof room, each participant provided two voice recordings by reading the Rainbow passage \cite{fairbanks1940voice} using a conversational tone. The choice of the Rainbow passage \cite{fairbanks1940voice} was to control for variation in linguistics and paralinguistics. As the Rainbow Passage \cite{fairbanks1940voice} is neutral in its content, it reduces the likelihood of evoking any pronounced emotional expression in the voice samples as well as standardizing linguistic features across all participants. Only the second sentence from the passage was used for the masculinity and femininity ratings.
\begin{table}[h!]
	\caption{Database description}
	\label{tab:method_preparations_datasets}
	\centering
	\begin{tabular}{llll}
		\toprule
		\textbf{Cohort No.}&\textbf{1}&\textbf{2}\\\midrule
		Collected year & 2015 & 2019\\\midrule
		Speakers mean age & 18.9 years&19.09 years\\\midrule
		Number of speakers& 22 M*, 22 F*&74 M, 107 F\\\midrule
		Number of raters&30&25\\\midrule
		Judgements&Mas$\ddagger$ for M, Fem$\ddagger$ for F&Mas for M, Fem for F\\\midrule
		Rating scale&1-10&1-100\\
		\bottomrule
		
		\multicolumn{3}{l}{* M - males; F - females}\\
		\multicolumn{3}{l}{$\ddagger$ Mas - masculinity; Fem - femininity}\\
	\end{tabular}
\end{table}

Human masculinity/femininity ratings were provided by Caucasian raters who did not know the speakers. For each rater, the voices were presented in two blocks (male voices and female voices), with block order counterbalanced across raters. Within each block, voices were presented in a random order. Following the presentation of each voice through enclosed headphones, a rating scale appeared on the screen. Listeners were asked to rate the masculinity for male speakers and to rate the femininity for female speakers, which is the same method used in \cite{cartei2014makes,little2011human,pisanski2011prioritization,munson2007acoustic,biemans2000gender} (see the “Judgement” column in Table \ref{tab:intro_relatedworks_summary}). The scale ranged from 1 to 10 for Cohort 1 and 1 to 100 for Cohort 2, with the extreme points labelled as ‘not at all masculine’ and ‘extremely masculine’ for male voices, and ‘not at all feminine’ and ‘extremely feminine’ for female voices. Note that, as illustrated in Table \ref{tab:intro_relatedworks_summary}, there is no universally agreed method for collecting judgements of perceived masculinity and femininity. The most popular methods are either to collect masculinity ratings for males and femininity ratings for females (as we have done) or to use a single masculine-feminine rating scale for both sexes. We did not use the single masculine-feminine rating scale method because of the evidence from Munson’s pilot study \cite{munson2007acoustic} that raters generally used only the masculine end of the scale for rating males and only the feminine end of the scale for rating females. Thus, we used a masculinity scale for males and a femininity scale for females to ensure a wide range of ratings within each set of voices. In addition, we did not apply both scales on each voice sample, because it would double the duration of the rating task; consequently, raters might be less engaged in the task, which could add noise to the rating data due to participants’ fatigue.

The recruitment and testing of all participants were conducted in accordance with the ethics approval obtained for this study from the Human Research Ethics Committee at the University of Western Australia.

\subsubsection{Acoustic Measures}
\label{txt:method_preparations_acousticmeasures}
The set of 23 widely known acoustic measures in Table \ref{tab:method_preparations_acousticmeasure} were used to describe the vocal characteristics of each speaker. Among these measures, mean value of F0 (F0 mean), standard deviation of F0 (F0 SD), HNR, all jitter measures (local Jitter, local absolute Jitter, rap Jitter, ppq5 Jitter and ddp Jitter)\footnote{http://www.fon.hum.uva.nl/praat/manual/Voice\textunderscore 2\textunderscore \textunderscore Jitter.html} and all shimmer measures (local Shimmer, apq3/5/11 Shimmer and dda Shimmer)\footnote{http://www.fon.hum.uva.nl/praat/manual/Voice\textunderscore 3\textunderscore \textunderscore Shimmer.html} were obtained from Parselmouth 0.3.3 which is a Python library for the Praat Software. The mean values of F1, F2, F3 and F4 measured the corresponding formants at each glottal pulse using the formant position formula \cite{puts2012masculine}. The apparent VTL was estimated in six measures: formant position (pF) \cite{puts2012masculine}, formant dispersion (fdisp) \cite{fitch1997vocal}, average formant frequency (avgFormant) \cite{pisanski2011prioritization}, geometric mean formant frequency (mff) \cite{smith2005interaction}, Fitch formant estimate \cite{fitch1997vocal} and formant spacing ($\Delta$F) \cite{reby2003anatomical}. 

\begin{table}[h!]
	\centering
	\caption{Acoustic measures}
	\label{tab:method_preparations_acousticmeasure}
	\resizebox{\textwidth}{!}{%
		\begin{tabular}{ll|ll|ll}
			\hline
			\multicolumn{2}{l|}{\textbf{Pitch related measures (\#1, \#2)}}                  & \multicolumn{2}{l|}{\textbf{Acoustic perturbation measure - shimmer (\#9 - \#13)}} & \multicolumn{2}{l}{\textbf{Vocal-tract length (VTL) estimates (\#18 - \#23)}} \\
			1                       & F0 mean                                     & 9                                  & local Shimmer                                 & 18                 & formant position (pF)                                    \\
			2                       & F0 standard deviation (F0 SD)                      & 10                                 & apq3 Shimmer                                  & 19                 & formant dispersion (fdisp)                               \\ \cline{1-2}
			\multicolumn{2}{l|}{\textbf{Acoustic perturbation measure - HNR (\#3)}}          & 11                                 & apq5 Shimmer                                  & 20                 & average formant frequency (avgFormant)                   \\
			3                       & HNR                                                    & 12                                 & apq11 Shimmer                                 & 21                 & geometric mean formant frequency (mff)                   \\ \cline{1-2}
			\multicolumn{2}{l|}{\textbf{Acoustic perturbation measure - jitter (\#4 - \#8)}} & 13                                 & dda Shimmer                                   & 22                 & fitch VTL                                                \\ \cline{3-4}
			4                       & local Jitter                                           & \multicolumn{2}{l|}{\textbf{Formant frequencies (\#14 - \#17)}}                    & 23                 & formant spacing ($\Delta F$)                \\
			5                       & local absolute Jitter                                  & 14                                 & F1 mean (F1)                                  &                    &                                                          \\
			6                       & rap Jitter                                             & 15                                 & F2 mean (F2)                                  &                    &                                                          \\
			7                       & ppq5 Jitter                                            & 16                                 & F3 mean (F3)                                  &                    &                                                          \\
			8                       & ddp Jitter                                             & 17                                 & F4 mean (F4)                                  &                    &                                                          \\ \hline
		\end{tabular}%
	}
\end{table}

\subsubsection{Extreme Random Forest Models}
\label{txt:method_preparations_ERF}
The Extreme Random Forest (ERF) is one of the most popular machine learning algorithms used for classification and regression purposes, providing good predictive performance, low over-fitting and easy interpretability \cite{geurts2006extremely}. In the case of regression, the ERF works by creating a large number of unpruned decision trees from the training dataset. Predictions are made by averaging the prediction of the decision trees. 

Furthermore, it is easy to obtain the contributions of each variable to the decision, by computing the impurity of each node. In regression mode, the measure of impurity is the variance. The principal idea is that the more a feature decreases the impurity, the more important the feature is. In the Random Forest (RF), the impurity decreases provided by each feature can be averaged across trees to determine the feature's importance. In other words, features that are selected at the top of trees are in general more important than features that are selected at the end nodes of trees, as top splits lead to bigger information gains. The ERF can be regarded as an extension of the RF \cite{breiman1984classification,breiman2001random,geurts2006extremely}. The main difference is that, the RF computes the locally optimal feature/split combination, while the ERF selects a random value for the split for each feature under consideration. Thus, the ERF uses more diversified trees and less splitters, so that the ERF is much faster than the RF with reduced tendency to overfit. 

In this study, the most suitable hyper-parameters of each ERF model were obtained by exhaustive search over specified parameter values and cross-validation splitting strategy of 4 folds, evaluated by the mean square error (MSE). Considering the hyper-parameters would vary with different types of input data, a set of ERF models were designed based on the input data size and the number of input data dimensions. The ERF models applied for the purpose of masculinity/femininity rating (ERF Model A in Figure \ref{fig:method_blockdiagram}) were trained on the 23 acoustic measures extracted from each segment from the input training data. The duration of the speech segments were varied from 1 second to 10 seconds to establish which duration yielded the best objective masculinity/femininity scoring performance (as described in Section \ref{txt:method_preprocessing_segmentation}). Then beginning with the 23 clusters (corresponding to the 23 acoustic measures), the number of clusters was progressively reduced down to just 1 cluster by the hierarchical clustering method described in Section \ref{txt:method_buildindepclusters}. These different numbers of clusters were then used to train a subset of ERF Models (ERF Model Sets B in Figure \ref{fig:method_blockdiagram}) to assess the quality of each cluster reduction. The optimal number of clusters was then determined as the minimum number of clusters with negligible reduction in performance (as described in Section \ref{txt:method_buildindepclusters_evaluation}). The optimal speech duration and the optimal number of clusters were then used for training a final ERF model (ERF Model C in Figure \ref{fig:method_blockdiagram}) to extract the feature importance for the acoustic factors characterization.

\subsection{Pre-processing}
\label{txt:method_preprocessing}
\subsubsection{Label Normalization}
\label{txt:method_preprocessing_labelnorm}
As mentioned above in Section \ref{txt:method_preparations_datasets} two different scales were applied in collecting perceived masculinity/femininity ratings, with the scales ranging from 1 to 10 for Cohort 1, and from 1 to 100 for Cohort 2. To correct for how listeners may have used the masculinity/femininity rating scales differently, the ratings provided by each listener were converted to z-scores. This also enabled the merging of ratings across the two cohorts. The label of each speaker was the mean value of all the z-scored perceived masculinity/femininity ratings given by the listeners.

\subsubsection{Segmentation and Acoustic Measures Extraction}
\label{txt:method_preprocessing_segmentation}
The audio files in the datasets were separated into the male and female sets which were used to build the ERF models for each gender independent of the other gender. Based on the literature from Table \ref{tab:intro_relatedworks_summary}, long-term speech segments were shown to be most suitable for studying the acoustic measures that affect perceived masculinity/femininity ratings, ranging from word-level utterances to utterances of multiple sentences. However, it is not evident what the best speech duration should be. In order to compare performances based on various speech durations, all the recordings were pre-processed to obtain the targeted speakers' utterances and were segmented into 1, 2, 5, 7 and 10 second speech duration datasets. Additionally, as the raters provided their perceived masculinity/femininity ratings only on the second sentence of the Rainbow passage (2 - 3 seconds in duration), the second sentences only were also regarded as one set of input data, which was compared with the various speech duration segments extracted from all the available utterances. 

Given the assumption that the perceived masculinity/femininity rating of one speaker would not vary across different utterances, all the segments provided by the same speaker shared the same label obtained as described in Section \ref{txt:method_preprocessing_labelnorm}.

For each input dataset (5 durations and the 2nd sentence), the set of 23 widely known acoustic measures (see Section \ref{txt:method_preparations_acousticmeasures}) were extracted for each segment. Each of the 23 acoustic measures were then z-normalised across each input dataset.

\subsection{Machine Rating of Masculinity and Femininity}
\label{txt:method_objectivegenderrating}
To investigate the optimal speech duration which achieves the best performance in masculinity/femininity rating, ERF model A (Figure \ref{fig:method_blockdiagram}) was trained on the 5 datasets with various speech durations and the 2nd sentence dataset. 

The utterances, provided by the speakers, were composed of two voice recordings of the Rainbow passage, with a total duration of $ N_i $ seconds for speaker $ i $. The utterances of each speaker were then segmented into $ L $-second segments and for each segment the 23 acoustic measures were extracted which are the input data samples. The input dataset details are specified in Table \ref{tab:method_objectivegenderraing_inputdata}. The number of samples was calculated using Equation \ref{eq:method_objectivegenderraing_nsamples}, where $ S $ denotes the set of female speakers and male speakers, respectively. The input data size is the number of samples $\times$ the 23 acoustic measures.

\begin{equation}
\label{eq:method_objectivegenderraing_nsamples}
\text{Number of Samples} = \sum_{i \in S} \lceil\frac{N_{i}}{L}\rceil
\end{equation}

\begin{table}[h!]
	\centering
	\caption{Number of samples for each segment duration across the 96 male and 129 female speakers}
	\label{tab:method_objectivegenderraing_inputdata}
	
		\begin{tabular}{lll}
			\hline
			\multirow{2}{*}{\textbf{Speech Duration $L$}} & \multicolumn{2}{l}{\textbf{Number of Samples}} \\
			& 96 Males         & 129 Females         \\ \hline
			1 second                         & 4341             & 6933               \\
			2 seconds                        & 2593             & 3720               \\
			5 seconds                        & 1167             & 1636               \\
			7 seconds                        & 871              & 1240               \\
			10 seconds                       & 676              & 957                \\
			2nd sentence                     & 96               & 129                \\ \hline
		\end{tabular}
\end{table}

To evaluate the performance of the ERF model, the standard k-fold cross validation was applied on each speaker. We used k = 4 and this was realised by partitioning the samples of each speaker into 4 folds, where 3 folds were used for training the ERF model (training data) and the remaining fold was used for validation (testing data). This procedure was then repeated 4 times so that each fold of one speaker was used once as the testing data. For the 2nd sentence input dataset however, we used 5-fold cross validation across all the samples.

The performance evaluation of the machine rating model of masculinity and femininity was carried out by investigating the mean values of the 4 sets of the $R^2$, $MSE$ and the Pearson correlation coefficient ($r$) values using either the training data or the testing data with the perceived masculinity/femininity rating labels. Specifically, within each fold iteration, the ERF model was trained on the training data, together with the corresponding perceived masculinity/femininity rating labels. The $R^2_{train}$, ${MSE}_{train}$ and $r_{train}$ were calculated based on the predictions of the training data, and the $R^2_{test}$, ${MSE}_{test}$ and $r_{test}$ were calculated based on predictions of the testing data both using the same trained ERF model. These procedures were repeated 4 times. The final $R^2_{train}$, $R^2_{test}$, ${MSE}_{train}$, ${MSE}_{test}$, $r_{train}$ and $r_{test}$ of each input dataset were obtained by calculating their mean values of the 4 sets.

The optimal speech durations for both genders were then determined based on the performances of the models. The final machine ratings of masculinity/femininity were generated by the model using the optimal speech duration. The details of the implementation are provided in \ref{app:pseudogenderrating}.

\subsection{Building Independent Clusters}
\label{txt:method_buildindepclusters}
Using all the samples with the optimal speech duration obtained from the previous step (Section \ref{txt:method_objectivegenderrating}), this section focuses on addressing the problem of multicollinearity and building independent clusters of acoustic measures.

\subsubsection{Correlation Matrix}

\begin{figure}[H]
	\centering
	\begin{subfigure}[b]{0.45\textwidth}
		\centering
		\includegraphics[width=1\linewidth]{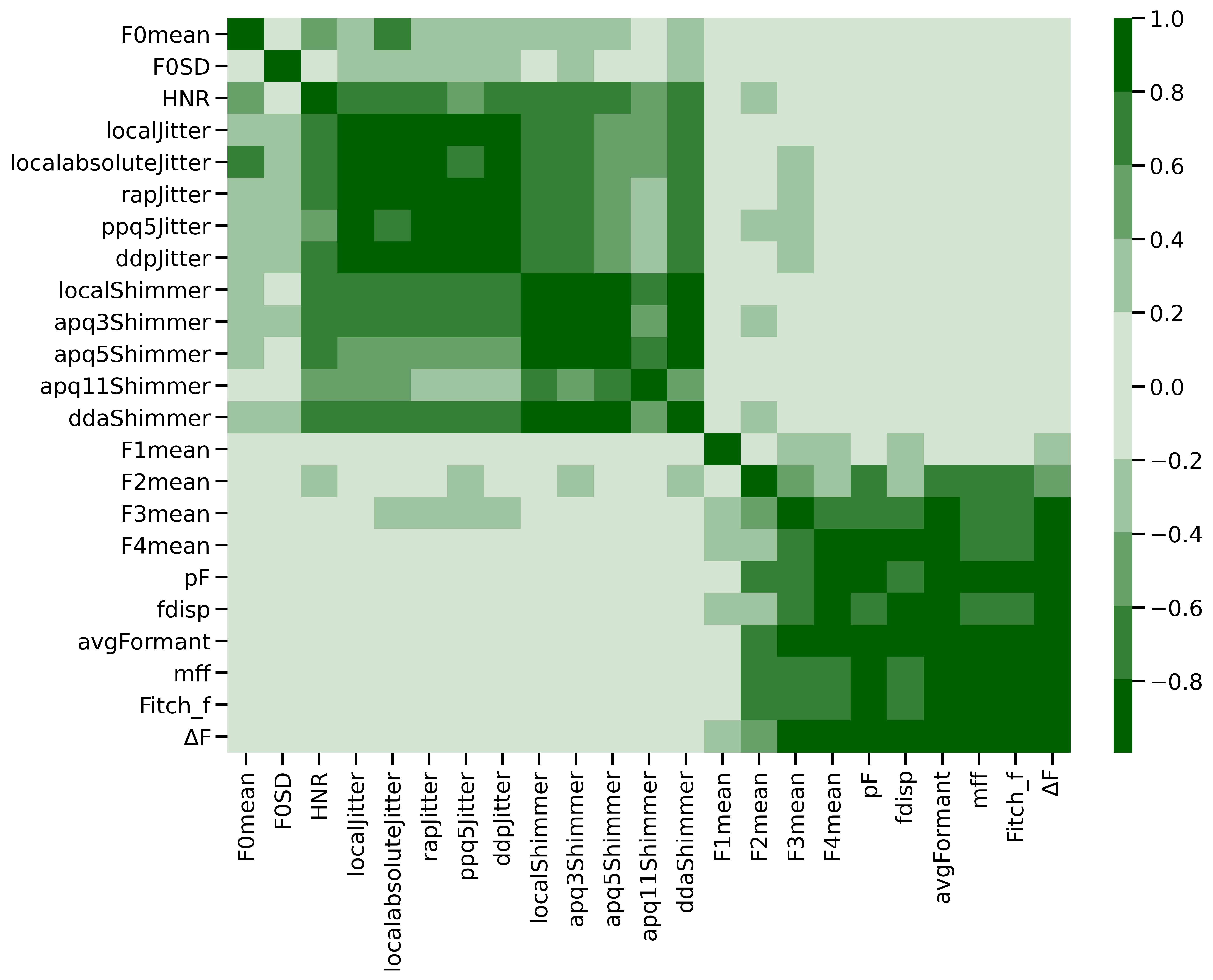}
	\end{subfigure}
	\qquad
	\begin{subfigure}[b]{0.45\textwidth}
		\centering
		\includegraphics[width=1\linewidth]{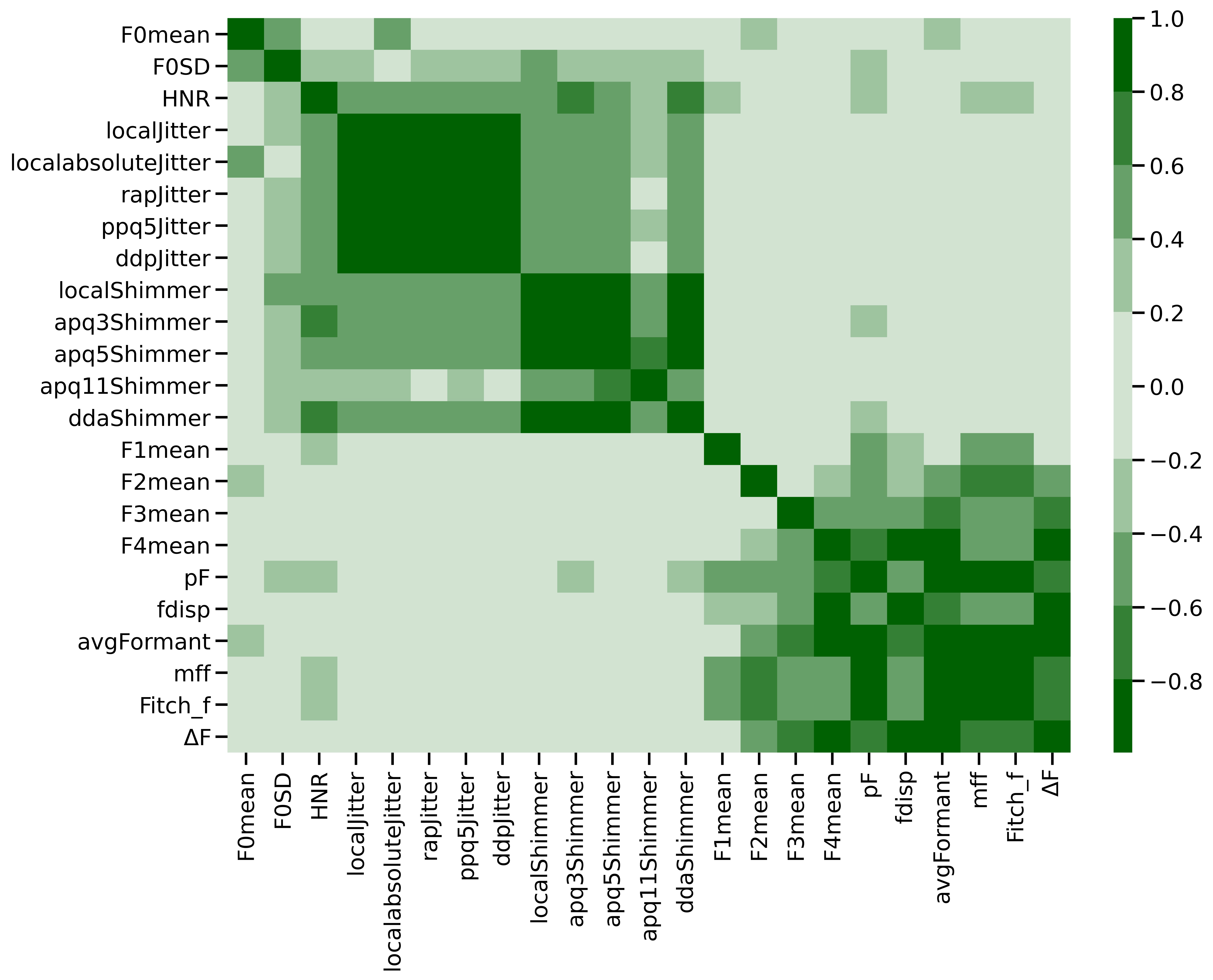}
	\end{subfigure}
	
	\caption{Correlation matrix of acoustic measures for females (left) and males (right), where the (u,v)th element is $r(u,v)$. The numerical values are provided in \ref{app:correlation}.}
	\label{fig:method_buildindepclusters_correlationmatrix_rawr}
\end{figure}

Figure \ref{fig:method_buildindepclusters_correlationmatrix_rawr} is a visual depiction of the correlation matrix for the pairs of acoustic measures, $ u $ and $ v $, calculated using Pearson's correlation coefficient $r(u,v)$. The correlation matrix is used to (1) apply hierarchical clustering and generate the dendrograms which will be discussed in Section \ref{txt:result_indepcluster_cluster}, and (2) supply an initial overview of the inter-correlations which will be compared with the final correlation matrix of independent clusters/acoustic measures. From Figure \ref{fig:method_buildindepclusters_correlationmatrix_rawr} it is evident that severe multicollinearity exists among the 23 acoustic measures, as some of the measures in the same group are highly correlated with each other, such as the measures in the jitter group, the shimmer group and the group of VTL estimators. Although the presence of multicollinearity is a common problem in both females and males, the severity may vary between females and males. Such differences may result in different clustering patterns.

\subsubsection{Hierarchical Clustering and Dendrogram}
\label{txt:method_buildindepclusters_cluster}
Correlation matrix based hierarchical clustering was proposed in study \cite{liu2012correlation} to identify the correlation structures among multiple features. We applied the same hierarchical clustering method to group the acoustic measures in clusters based on the acoustic measures' similarity. The idea is to regard each acoustic measure as an individual cluster and then merge them into the nearest clusters, until one cluster remains. The hierarchical clustering approach achieves this by generating a dendrogram, which is a tree-based representation of the 23 acoustic measures. The iterations (shown as the loop in Figure \ref{fig:method_blockdiagram}), including multicollinearity monitoring, clusters representation and training the ERF Model Sets B, were carried out based on the dendrogram. The method of generating the dendrogram is as follows.

Firstly, the similarities for all pairs of the acoustic measures were measured, by investigating the correlation matrix of these 23 acoustic measures (see Figure \ref{fig:method_buildindepclusters_correlationmatrix_rawr}). Before clustering, we needed to define the distance between two acoustic measures. Being different from study \cite{liu2012correlation}, that measured the distance by $ d(u,v) = 1 - r(u,v) $, the present study considered another distance measure. The distance $ d(u,v) $, measuring the similarity between the two acoustic measures $ u $ and $ v $, was defined by the Euclidean distance between all the points of $ u $ and $ v $, defined by Eq \ref{eq:method_buildindepclusters_cluster_duv}. The more similar the two acoustic measures are, the shorter the distance is. In study \cite{gu2010study}, it was shown such a distance measure performed better than the one used in \cite{liu2012correlation} in hierarchical clustering tasks. It is because the distance measure $ d(u,v) = 1 - r(u,v) $ only considered the correlation between the two acoustic measures, whereas by using the distance measure Eq \ref{eq:method_buildindepclusters_cluster_duv}, we considered all the correlations of the targeted acoustic measure with the other 22 acoustic measures.
\begin{equation}
\label{eq:method_buildindepclusters_cluster_duv}
d(u,v)= \sqrt{\sum_{k=1}^{23}(|r(u,k)|-|r(v,k)|)^2}
\end{equation}

More generally, it is necessary to introduce a distance measurement called cophenetic distance, $ D(s,t) $, to measure the distance between any two clusters $ s $ and $ t $ which could contain multiple acoustic measures. There are different ways to calculate the $ D(s,t) $; among them the single, complete, average and centroid linkages have been the most commonly used. In this study, the average linkage method was used to compute the $D(s,t)$ between any two clusters $ s $ and $ t $, which is defined by Eq \ref{eq:method_buildindepclusters_cluster_average}, where $ u $ and $ v $ denote all the points in the clusters $ s $ and $ t $, and $ N_s $ and $ N_t $ are the cardinalities of the clusters $ s $ and $ t $, respectively. The average linkage was chosen because it yielded higher cophenetic correlations than the other methods. Besides, it reduced the tendency to produce chain-shape clusters which always occurs by using single linkage, and average linkage has a higher tolerance for outliers than the complete linkage \cite{liu2012correlation}.
\begin{equation}
\label{eq:method_buildindepclusters_cluster_average}
D(s,t)=\frac{\sum_{u=1}^{N_s}\sum_{v=1}^{N_t}d(u,v)}{N_s\cdot N_t}
\end{equation}

The flow of the hierarchical clustering started by finding the shortest distance $ d_{min}(u,v) $ across all pair combinations $ \{u,v\} $ of the 23 initial acoustic measures and merging them into the first cluster $ c \{u,v\} $. This resulted in 22 clusters: cluster $ c \{u,v\} $ and 21 singleton clusters of the remaining, yet to be merged, acoustic measures. Eq \ref{eq:method_buildindepclusters_cluster_average} was used to calculate each pair-wise cophenetic distance $ D $ among the 22 acoustic measures to find the clusters $ s $ and $ t $ with the minimum distance $ D_{min}(s,t) $ which were then merged into a new cluster. The clustering stopped when all the 23 acoustic measures formed into one cluster. The pseudo-code implementation is provided in \ref{app:pseudocluster}.

The above procedure generates the dendrogram which will be discussed in Section \ref{txt:result_indepcluster_cluster}. The dendrogram provides the essential information for the subsequent iterations. The information conveyed by the dendrogram includes the acoustic measures clustering order, the clustering patterns and the cophenetic distances.

\subsubsection{Multicollinearity Monitoring}
\label{txt:method_buildindepclusters_multicolmonitor}
The severity of multicollinearity was monitored throughout the iterations, which supplies important information for the determination of the optimal number of clusters. 

The multicollinearity issue would not significantly affect the prediction of the ERF Model Sets B (see Figure \ref{fig:method_blockdiagram}), but would strongly and negatively affect the identification of feature importance when using the ERF model. As mentioned above, in the application of the ERF, each tree of the ERF would pick the most discriminant variable which owns the lowest impurity, while the other correlated variables would be less important in accounting for any further variation in masculinity/femininity ratings. Across a large number of trees in the ERF, the overall importance of any highly correlated variables would be reduced. Therefore, addressing the multicollinearity problem is essential for meaningful acoustic characterization using the ERF model.

The Variance Inflation Factor (VIF) is the quotient of the variance in a model with multiple terms by the variance of a model with one term alone \cite{james2013introduction}, which has been popularly used to assess the severity of multicollinearity among the measures and clusters. In this study, assuming there are $ k $ variables, the data is the correlation matrix of $ k $ variables, with the size of $ k \times k $ ($x_1$,$x_2$, ..., $x_k$, $1\leq k\leq 23$). Firstly an ordinary least square regression was applied on each variable $x_i$
\begin{equation}
\label{eq:method_buildindepclusters_multicolmonitor_xi}
x_i = a_0+a_1x_1+...+a_{i-1}x_{i-1}+a_{i+1}x_{i+1}+...+a_kx_k+e
\end{equation}
where $a_0$ is a constant and $ e $ is the error term. Secondly, each VIF value $VIF_i$ corresponding to $x_i$ was calculated as
\begin{equation}
\label{eq:method_buildindepclusters_multicolmonitor_vif}
VIF_i=1/(1-R_i^2)
\end{equation}
where $R_i^2$ is the coefficient of determination of the regression equation. We analysed the magnitude of multicollinearity by considering the size of the VIFs. A rule of thumb is that if the largest VIF is greater than 5, then multicollinearity is high \cite{craney2002model,sheather2009modern}.

Initially before clustering, a VIF was calculated for each of the 23 acoustic measures ($k$ = 23), with the VIF representing the severity of multicollinearity of each measure with the other measures. When the clustering iteration began, the number of clusters, k, was progressively reduced until it became 1. Within each iteration, the VIFs were recalculated based on the updated correlation matrix of all clusters. For example, after the $i_{th}$ iteration, the acoustic measure variables were grouped into $ k $ clusters. Among the $ k $ clusters, $k_m$ clusters were composed of multiple acoustic measures, while the remaining ($k-k_m$) clusters were the initial single acoustic measures. The multiple acoustic measures in each $k_m$ cluster were considered to be correlated, and were replaced by a single variable using Principal Component Analysis (PCA) (see Section \ref{txt:method_buildindepclusters_clusterrepresent}). In the next iteration, these newly generated variables for each of the $ k_m $ clusters, together with the $k-k_m$ single measures, were used to update the correlation matrix, with the size of $ k \times k $, as well as the latest VIFs.

\subsubsection{Cluster Representation}
\label{txt:method_buildindepclusters_clusterrepresent}
According to the dendrogram generated from Section \ref{txt:method_buildindepclusters_cluster}, the iteration started at the first joint point defined as where the two most correlated acoustic measures were merged, and ended up at the last joint point where all acoustic measures were merged into one cluster. 

For each cluster among the $ k_m $ clusters, a PCA model was applied to generate one principal component to represent the cluster, by creating one new variable that miximized the variance of the enclosed values of acoustic measures. These newly generated variables, together with the $ k-k_m $ single acoustic measures, yielded a $ k $-dimensional vector $ S_{new} $, which was used to replace the sample $ S $ before clustering. By applying this procedure on all the samples, the newly generated $ S_{new} $ were then used to train the ERF Model Sets B (see Figure \ref{fig:method_blockdiagram}). 

\subsubsection{Evaluation}
\label{txt:method_buildindepclusters_evaluation}
The same data splitting strategy and model performance indicators as used in Section \ref{txt:method_objectivegenderrating} were used for the purpose of building independent clusters. The evaluation in this section concerned assessing the following three factors as the number of clusters were reduced from 23 to 1: (1) severity of multicollinearity, (2) system performances, and (3) clearly interpretable meanings of clusters. The optimal clusters were then obtained based on the assessments of these three factors.

Firstly, the severity of multicollinearity was monitored by investigating the VIFs within each iteration. It is expected that the VIFs would be extremely high at the beginning of clustering, according to the correlation matrix shown in Figure \ref{fig:method_buildindepclusters_correlationmatrix_rawr}. But the severity of multicollinearity would be relieved when some of the highly correlated measures were clustered and represented by the single variables using the PCA models. The absence of severe multicollinearity was only considered to be achieved when the maximum VIF value was lower than 5 \cite{craney2002model,sheather2009modern}.

Secondly, the six performance indicators from Section \ref{txt:method_objectivegenderrating}, $R^2_{train}$, $R^2_{test}$ , ${MSE}_{train}$, ${MSE}_{test}$, $r_{train}$ and $r_{test}$, were recorded at each iteration, for the training data and the testing data respectively. The best system performance is expected to happen before any clustering, because no information will be lost at this stage. The performance should remain unchanged or only degrade slightly before any critical independent acoustic measures (such as F0 and F1) are clustered into other unrelated groups. 

Thirdly, the interpretation of clusters was evaluated by investigating the enclosed acoustic measures of each cluster. At the beginning of clustering, as the most correlated acoustic measures are grouped, the physical meanings are supposed to be very explicit and interpretable, such as the cluster composed of mff and Fitch\_f (r = .99 for both males and females as shown in Figure \ref{fig:method_buildindepclusters_correlationmatrix_rawr} and \ref{app:correlation}) which are the estimators of the physical vocal-tract length (VTL). With larger and fewer clusters, the interpretation of each cluster is less clear. According to the current literature, the final determined optimal clusters should confirm that: (1) F0 mean and VTL estimators should appear in two independent clusters \cite{feinberg2006menstrual,feinberg2005manipulations}; (2) F0 mean and Fn mean should be in different clusters \cite{pisanski2011prioritization,fant1970acoustic,muller1848physiology}; and (3) 5 jitter measures should be grouped in one cluster, as similarly should the 4 shimmer measures.

As stated above, our method of building independent clusters of acoustic measures aims to eliminate multicollinearity while preserving the best system performance and retaining the clearly interpretable meanings of the clusters. The optimal number of clusters was determined by assessing the above three aspects and was used for the characterisation purpose. The criteria for the judgement of the optimal number of clusters include:

\begin{itemize}
	\item No severe multicollinearity exists, which means all the VIF values of these clusters must be lower than 5.
	\item The system performance must be no worse than the best performance happening before clustering.
	\item The physical meanings of each cluster must be easy to interpret with no obscureness.
\end{itemize}

\subsection{Characterization}
\label{txt:method_characterization}
In the characterization the best speech duration (from Section \ref{txt:method_objectivegenderrating}) and the optimal number of clusters (from Section \ref{txt:method_buildindepclusters}) were then used to train the ERF model C (see Figure \ref{fig:method_blockdiagram}) and generate the cluster weights to determine the feature importance of each cluster. The ERF model C was fined-tuned by using the most suitable hyper-parameters for the clustered data with the dimensions of the number of optimal clusters, as described in Section \ref{txt:method_buildindepclusters_evaluation}. The results of characterization were evaluated by comparing with what is known from the literature, summarised in Table \ref{tab:intro_relatedworks_summary}. The details of the implementation of building independent clusters (Section \ref{txt:method_buildindepclusters}) and characterization are provided in \ref{app:character}.

\section{Results and Discussion}
\subsection{Machine Rating of Masculinity and Femininity}
The best performance of each input dataset occurs prior to clustering. This is because the ERF will explore all the information across the full dimension range, whereas the clustering together with the PCA representation reduces the dimensions which results in missing some information. Table \ref{tab:result_objectivegenderrating_speechduration} shows the system performance of using different speech durations (1-, 2-, 5-, 7- and 10-second durations). As the human raters provided their ratings on masculinity and femininity based on listening to only the second sentence of speech for each speaker, the result of using the second sentence speech is also provided in Table \ref{tab:result_objectivegenderrating_speechduration} for comparison. The hyper-parameters of all the ERF models were fine-tuned to minimise model over-fitting.

\begin{table}[H]
	\caption{System performance of using different speech durations in females and males}
	\label{tab:result_objectivegenderrating_speechduration}
	\centering
	\resizebox{\columnwidth}{!}{
	\begin{tabular}{c|ccccccc|ccccccc}
		\hline
		\multicolumn{1}{l|}{}        & \multicolumn{7}{c|}{\textbf{Female}}                                                                        & \multicolumn{6}{c}{\textbf{Male}}                                                         &                 \\ \hline
		\textbf{Time frame (second)} & \textbf{1} & \textbf{2} & \textbf{5} & \textbf{7}   & \textbf{10} & \textbf{2nd sentence} & \textbf{7 (LR)} & \textbf{1} & \textbf{2} & \textbf{5} & \textbf{7}   & \textbf{10} & \textbf{2nd sentence} & \textbf{7 (LR)} \\ \hline
		\rowcolor[HTML]{C0C0C0} 
		$R^2_{train}$                           & .43        & .52        & .61        & \textbf{.73} & .55         & \textbf{.68}          & \textbf{.30}    & .65        & .77        & .53        & \textbf{.78} & .50         & \textbf{.84}          & \textbf{.50}    \\
		$R^2_{test}$                      & .20        & .27        & .35        & \textbf{.37} & .35         & \textbf{.19}          & \textbf{.26}    & .40        & .48        & .47        & \textbf{.57} & .44         & \textbf{.35}          & \textbf{.46}    \\
		\rowcolor[HTML]{C0C0C0} 
		$MSE_{train}$                   & .20        & .17        & .14        & \textbf{.09} & .16         & \textbf{.11}          & \textbf{.24}    & .21        & .13        & .28        & \textbf{.13} & .30         & \textbf{.09}          & \textbf{.29}    \\
		$MSE_{test}$                    & .27        & .25        & .23        & \textbf{.22} & .23         & \textbf{.28}          & \textbf{.26}    & .35        & .31        & .32        & \textbf{.25} & .33         & \textbf{.37}          & \textbf{.31}    \\
		\rowcolor[HTML]{C0C0C0} 
		$r_{train}$           & .72        & .79        & .84        & \textbf{.91} & .81         & \textbf{.90}          & \textbf{.55}    & .83        & .91        & .75        & \textbf{.90} & .74         & \textbf{.95}          & \textbf{.70}    \\
		$r_{test}$            & .46        & .54        & .61        & \textbf{.63} & .62         & \textbf{.46}          & \textbf{.51}    & .64        & .70        & .70        & \textbf{.77} & .68         & \textbf{.63}          & \textbf{.68}    \\ \hline
	\end{tabular}}
\end{table}

The analyses showed that model performance was better for the speech duration of 7 seconds compared to the other durations, in both genders. This outcome is consistent with the findings of previous studies \cite{cartei2014makes,harb2005voice}, that long-term speech is better than short-term speech in investigating the relationship between perceived masculinity/femininity ratings and features extracted from acoustic signal analysis. In using just binary labels for biological sex, our previous work \cite{chen2020objective} demonstrated that speakers' acoustic-related masculinity and femininity could be rated by an LDA model ($ r $ = .51 in females and $ r $ = .67 in males). In contrast, the current work using an ERF regression model to predict speakers' acoustic-related masculinity and femininity demonstrated a more promising result, with $ r_{test} $ = .63 in females and $ r_{test} $ = .77 in males.

As shown in Table \ref{tab:result_objectivegenderrating_speechduration}, when the 2nd sentence speech data were used, there was a large discrepancy in system performance between training data and testing data. The discrepancy indicates that even though the ERF model was correctly trained, the model exhibited over-fitting when the 2nd sentence speech data were used. One possible explanation could be that the size of the 2nd sentence speech data is much smaller than the other input datasets with different speech durations. Actually, each speaker provided 2 recordings of the same passage of approximately 30 seconds duration each, while only the speech of the 2nd sentence selected from one recording was rated by the human listeners. The speech duration of the rated 2nd sentence was only about 3 seconds, which was approximately 1/20th of the entire speech duration across the two recordings.

Nevertheless, the current results show discrepancy between the results of the training data and the testing data in using the data of 7 seconds duration, with $R^2_{train}$=.73 and $R^2_{test}$=.37 for females, and $R^2_{train}$=.78 and $R^2_{test}$=.57 for males. Therefore, we evaluated the model for possible over-fitting by comparing the ERF model results from those obtained from a Linear Regression (LR) model, using 7-second duration data for training and testing, and 4-fold cross validation for both models. The results demonstrated that the performance of using the 7-second duration data (see column `7' in Table \ref{tab:result_objectivegenderrating_speechduration}) is better than the results based on LR (see column `7 (LR)' in Table \ref{tab:result_objectivegenderrating_speechduration}), for both the training and testing data, even though there is less over-fitting using the LR (e.g. $R^2_{train}$ = .30 and $R^2_{test}$ = .26)

We present a possible explanation for the discrepancy between the results of the training data and the testing data in the ERF model using the 7-second speech duration data. The perceived masculinity/femininity ratings were provided based on only the second sentence of the speech (almost 3 seconds), which was about 5\% of the entire speech length for each speaker (almost 60 seconds). The mean value of the perceived masculinity/femininity ratings for this speech segment was used to label all the speech segments from each speaker and was regarded as the “ground truth”. By using this generalization method, we were not limited to using only the second sentence (i.e. 225 samples each 2-3 seconds in duration) in model training and testing. Instead, a much larger dataset of speech samples could be used (e.g. 2111 samples for the 7-second duration), which significantly benefited the ERF model training. However, the side effect was that all the segments of one speaker shared one perceived masculinity/femininity rating. Variation may exist throughout a 30 second recording which may influence the masculinity/femininity rating for any 2-3 second segment. In the present study, the ERF model was given a collection of samples from each speaker with different values of acoustic measures expressed across the samples, but was forced to regress these different samples on to one specific masculinity/femininity rating. This limited the model to precisely predict the rating for an unknown sample, which may have contributed to the gap between the training data and testing data results.

Also of note is the difference in model performance for males and females, with $r_{test}$ = .63 in females while $r_{test}$ = .77 in males for the 7-second duration data. This suggests that the ERF model tends to predict masculinity ratings for males better than femininity ratings for females. One possible reason for the difference in $R^2$, $MSE$ and $r$ in males and females is that human raters in this study were better at using vocal cues in judging masculinity in males than using vocal cues in judging femininity in females. Lippa \cite{lippa1978naive} reported a stronger correlation between observers' masculinity ratings of speakers' voices and speakers' masculinity in terms of their personalities for males (r = .59) compared to females (r = .49). In contrast, Lippa reported a stronger correlation between observers' masculinity ratings of speakers' faces (presented without voice) and speakers' masculinity in personality styles for females (r = .64) compared to males (r = .39). These results suggest that observers relied on vocal cues in making judgements of masculinity in males and relied on visual cues in making femininity judgements in females. Similarly in the study of facial gender scoring \cite{gilani2014geometric}, the machine ratings were more highly correlated with human perceived ratings in females (r = .90) than in males (r = .79). The results of the present study are consistent with the former literatures in the way that machine learning models, using acoustic measures as vocal cues, are more powerful in predicting masculinity for males than in predicting femininity for females.

In summary, our model enables prediction of the perceived masculinity or femininity for an unknown speaker based on a speech sample from the speaker. It is also worthwhile to mention that the predictions of the speakers' masculinity or femininity were derived from a data-driven model that was trained on a set of read speech samples and a set of human ratings which were collected from a specific group of participants and raters. This means the prediction of the perceived masculinity or femininity of an unknown speaker would be obtained from an algorithm instead of individuals. The performance of the model would be expected to be at its highest for Caucasian speakers and for a speech sample of read text that yields scores on the acoustic measures within the range of scores in the dataset used to train the model. The prediction could vary from the ratings provided by individuals, if the acoustic measures obtained from the speaker's speech are out of the range from our dataset. It could also vary from the human perceived ratings depending on the ethnicity, gender identity, sexual orientation, age and the type of speech of the unknown speaker. 

\subsection{Building Independent Clusters}
Using all the samples with the optimal speech duration of 7 seconds, this section provides the results and analysis of the hierarchical clustering and the determined independent clusters.

\subsubsection{Hierarchical Clustering and Dendrogram}
\label{txt:result_indepcluster_cluster}
\begin{figure}[H]
	\centering
	\begin{subfigure}[b]{0.45\textwidth}
		\centering
		\includegraphics[width=1\linewidth]{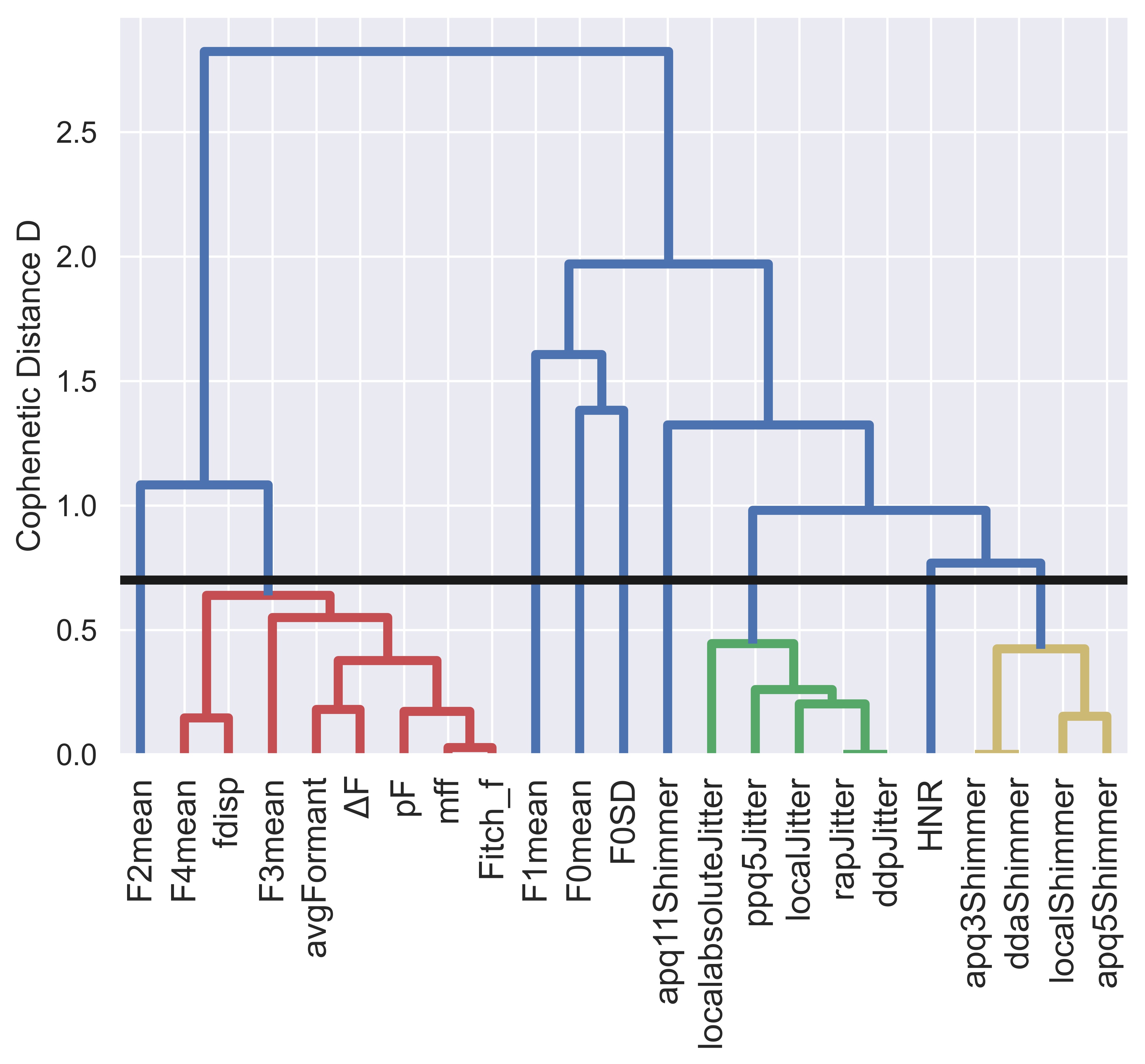}
	\end{subfigure}
	\qquad
	\begin{subfigure}[b]{0.45\textwidth}
		\centering
		\includegraphics[width=1\linewidth]{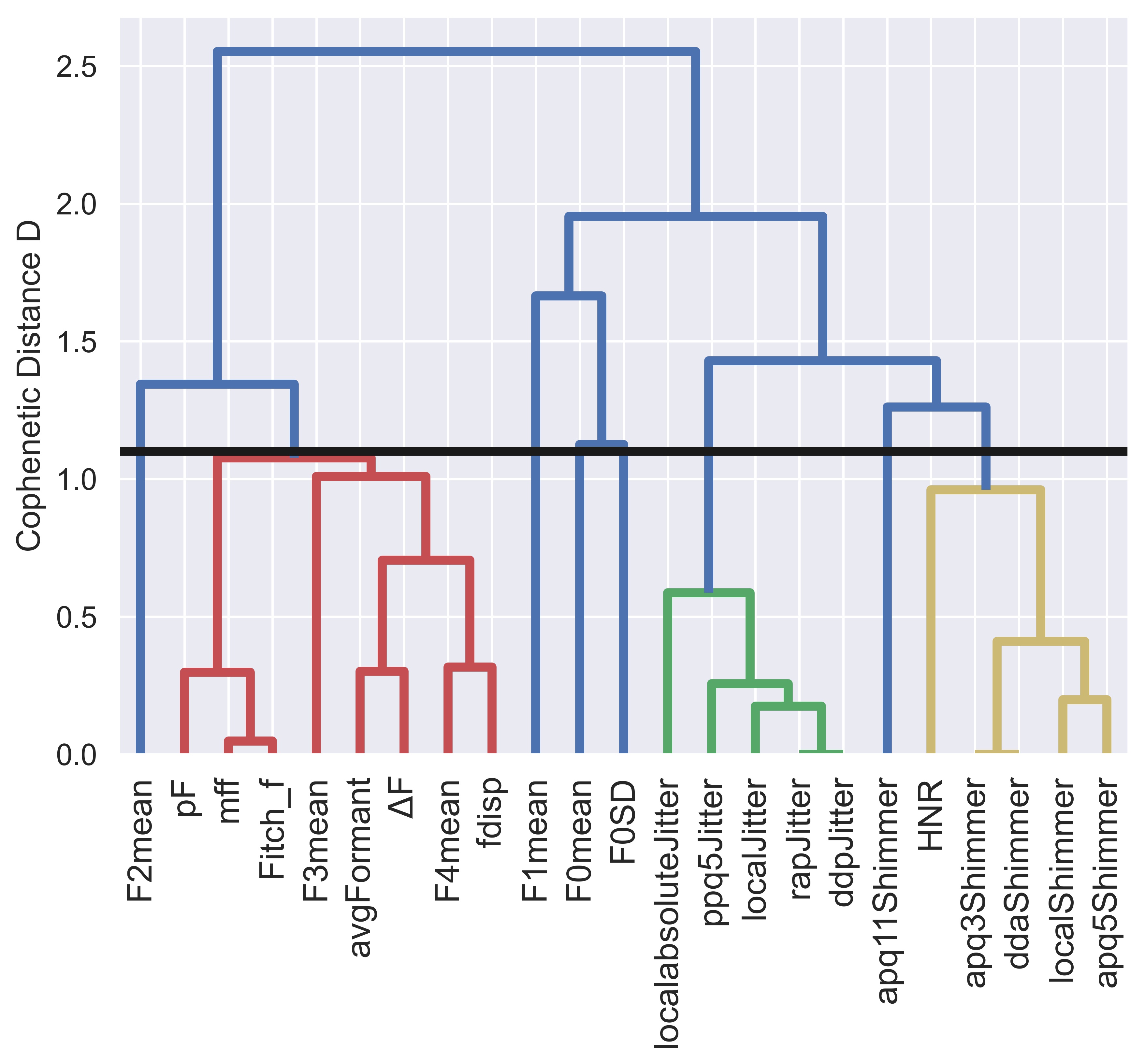}
	\end{subfigure}
	
	\caption{Dendrograms of acoustic measures in females (left) and males (right)}
	\label{fig:result_indepcluster_cluster_dendro}
\end{figure}
Figure \ref{fig:result_indepcluster_cluster_dendro} shows the dendrograms of the hierarchical clustering in the data of females and males. The common patterns shown in both males and females are: (1) all jitter-related measures (acoustic measures \#4 - \#8 in Table \ref{tab:method_preparations_acousticmeasure}) were formed into one cluster at an early stage in both females and males (cophenetic distance D $=$ .43 in females and D $=$ .59 in males, shown in green in Figure \ref{fig:result_indepcluster_cluster_dendro}), (2) all shimmer-related measures except for apq11shimmer (acoustic measures \#9, \#10, \#11 and \#13 in Table \ref{tab:method_preparations_acousticmeasure}) were clustered into one group in both males and females (D $=$ .42 in females and D $=$ .41 in males, shown in yellow in Figure \ref{fig:result_indepcluster_cluster_dendro}), (3) F4 mean and fdisp were highly correlated and were formed into one cluster (D $=$ .15 in females and D $=$ .32 in males, shown in red in Figure \ref{fig:result_indepcluster_cluster_dendro}), and (4) each single measure of F0 mean, F0 SD, F1 mean and F2 mean had D over 1 when merged with other clusters and measures (shown as the leaves of the blue branches in Figure \ref{fig:result_indepcluster_cluster_dendro}) and so were retained as individual measures.



\subsubsection{Minimum Multicollinearity Clustering}
\label{txt:result_indepcluster_multicol}

\begin{figure}[ht]
	\centering
	\begin{subfigure}[b]{0.49\textwidth}
		\centering
		\includegraphics[width=1\textwidth]{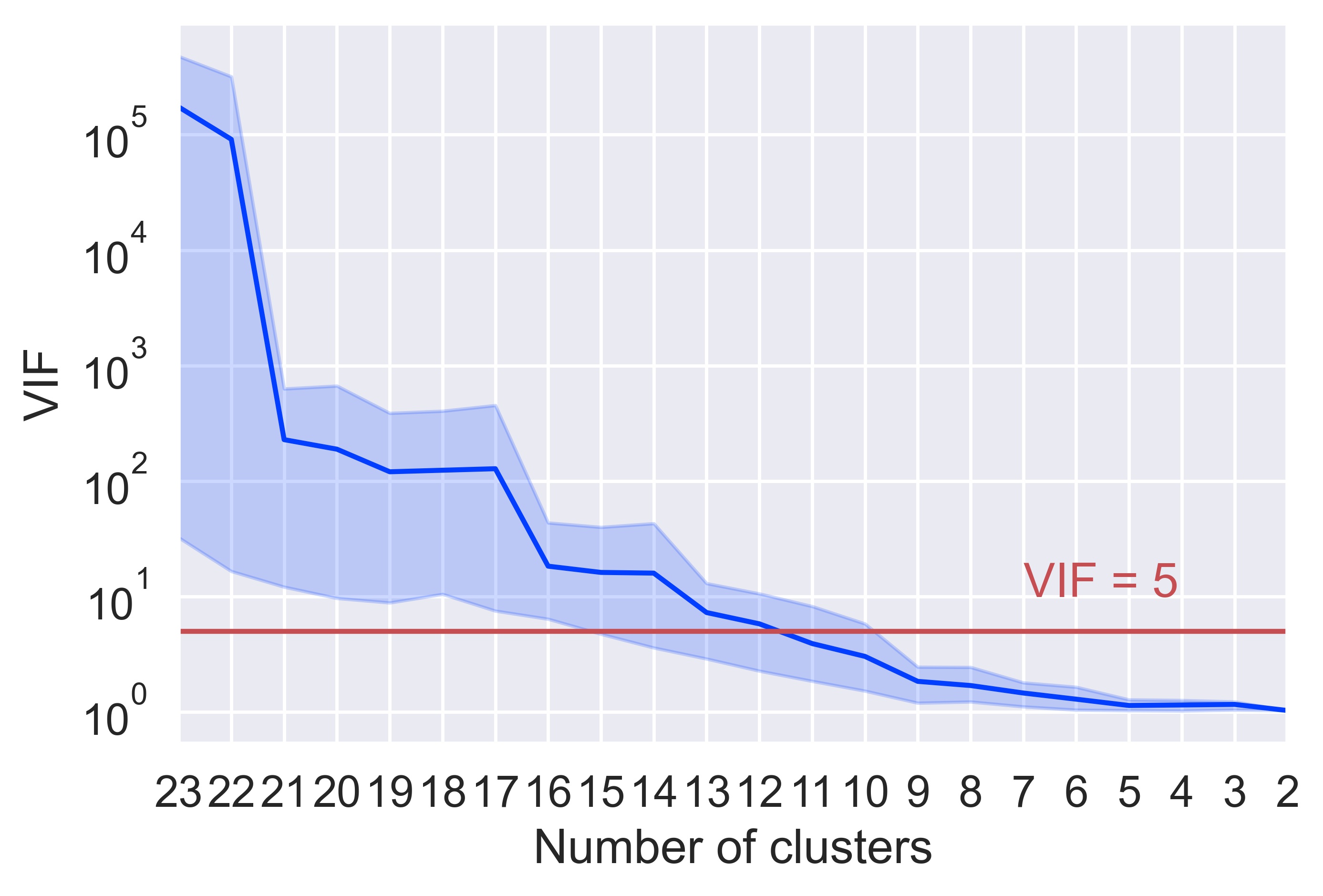}
	\end{subfigure}
	\begin{subfigure}[b]{0.49\textwidth}
		\centering
		\includegraphics[width=1\textwidth]{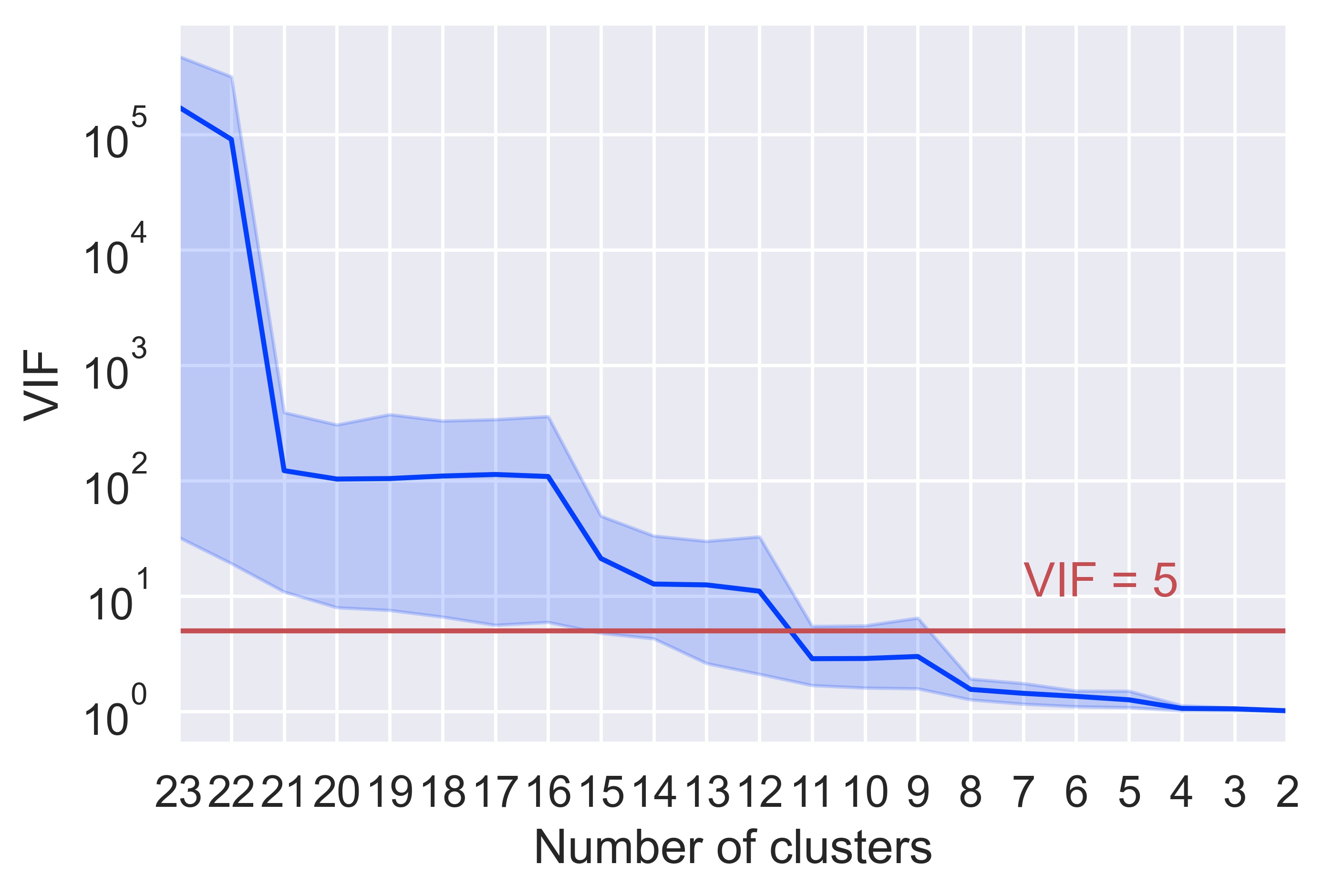}
	\end{subfigure}
	\caption{Relationships between VIFs and the number of clusters in females (left) and males (right)}
	\label{fig:result_indepcluster_multicol_vif}
\end{figure}

As mentioned in Section \ref{txt:method_buildindepclusters_multicolmonitor} it is only when the maximum value of VIFs is below 5 that the risk of multicollinearity is not considered a serious problem when using regression algorithms. Figure \ref{fig:result_indepcluster_multicol_vif} presents the VIF values with the number of clusters decreasing gradually, where the solid blue line demonstrates the mean values of the VIF values, the blue ribbon illustrates the range of the VIF values and the red horizontal line indicates the threshold of the VIF values which is set to 5. It is not surprising that the VIF values decreased more sharply at the beginning of clustering (as the highly correlated acoustic measures were grouped first) and converged more gradually to a low value at the end. For females, the maximum VIF value was below 5 when the number of clusters is at most 9 and for males at most 8. This yielded \textbf{9 uncorrelated clusters for females} and \textbf{8 uncorrelated clusters for males} as the meaningful acoustic factors.

\begin{figure}[H]
	\centering
	\begin{subfigure}[b]{0.49\textwidth}
		\centering
		\includegraphics[width=1\linewidth]{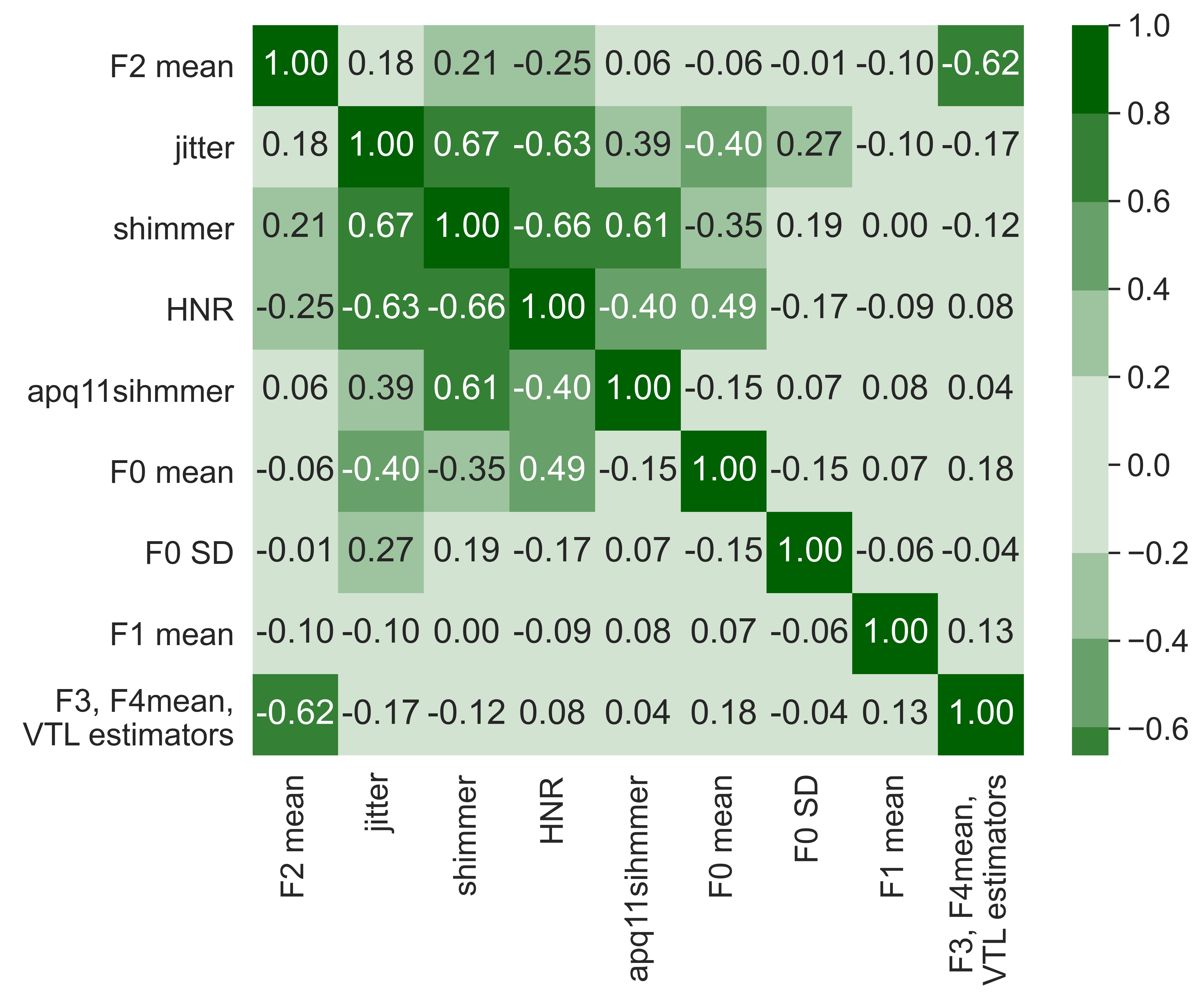}
	\end{subfigure}
	\begin{subfigure}[b]{0.49\textwidth}
		\centering
		\includegraphics[width=1\linewidth]{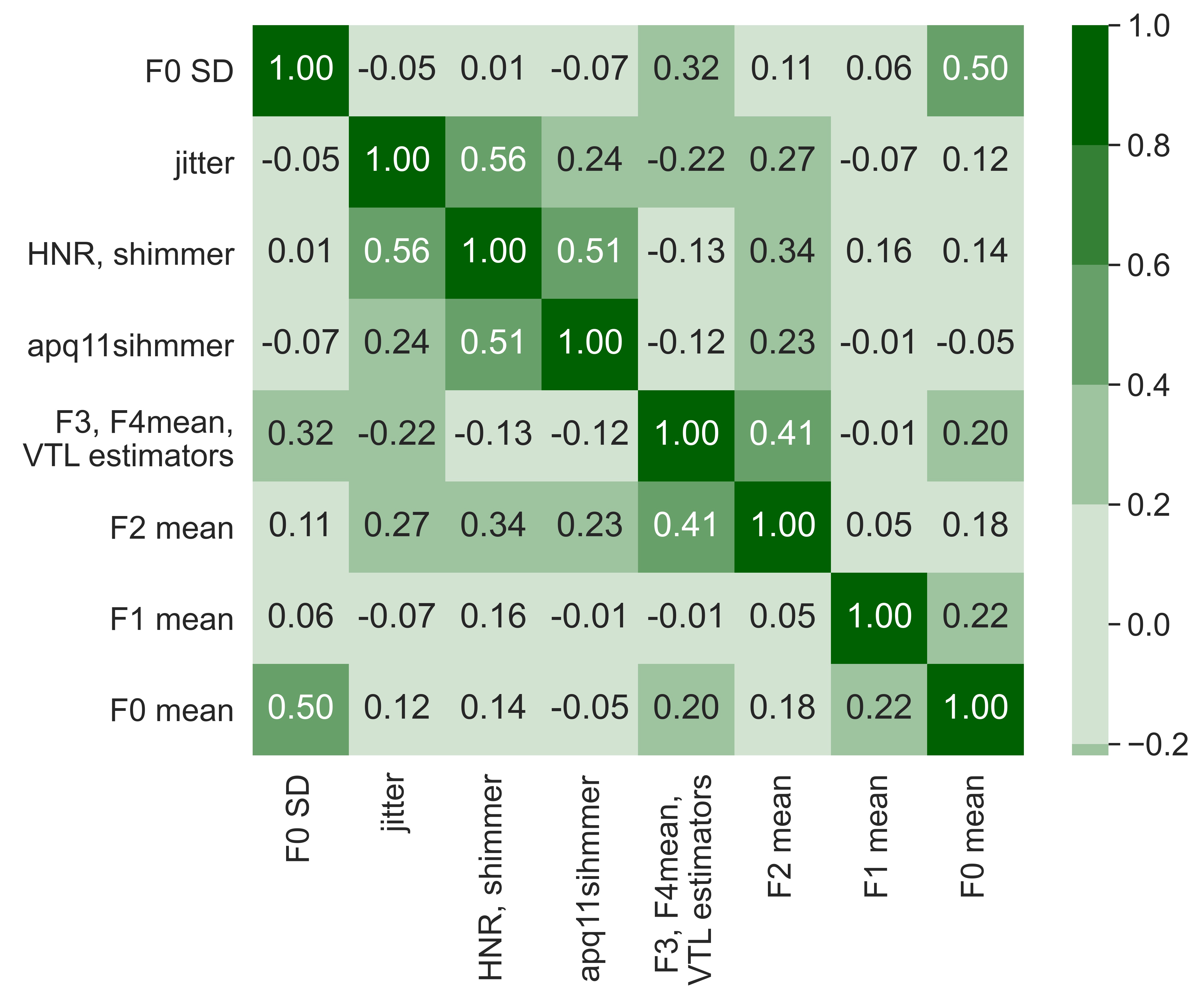}
	\end{subfigure}
	
	\caption{Correlation matrix of clustered acoustic measures in females (left) and males (right)}
	\label{fig:result_indepcluster_multicol_newr}
\end{figure}

The PCA was applied on each cluster to generate a new variable which represented the cluster. The correlation matrix of the clustered acoustic measures with 9 clusters for females and 8 clusters for males are shown in Figure \ref{fig:result_indepcluster_multicol_newr}. Table \ref{tab:result_indepcluster_multicol_cluster} shows each cluster, enclosed acoustic measures and VIF values for females and males. Comparing Figure \ref{fig:result_indepcluster_multicol_newr} with the correlation matrix before clustering (illustrated in Figure \ref{fig:method_buildindepclusters_correlationmatrix_rawr} and \ref{app:correlation}), Figure \ref{fig:result_indepcluster_multicol_newr} shows that no severe multicollinearity existed after clustering 23 acoustic measures into 9 independent clusters for females and 8 independent cluster for males. It should be noted that the highest inter-correlation, appearing in the three clusters in females (see Figure \ref{fig:result_indepcluster_multicol_newr}), namely jitter measures, shimmer measures and HNR, also yielded the highest VIF values after clustering, but these were all less than 5 (see Table \ref{tab:result_indepcluster_multicol_cluster}).

\begin{table}[H]
	\caption{Enclosed acoustic measures and VIFs of the independent clusters for females (left) and males (right)}
	\label{tab:result_indepcluster_multicol_cluster}
	\centering
	\resizebox{\columnwidth}{!}{
		\begin{tabular}{p{0.5\linewidth}p{0.5\linewidth}}
			\begin{tabular}{|p{0.35\columnwidth}|p{0.05\columnwidth}|}
				\hline
				\multicolumn{2}{|c|}{Female}                     \\ \hline
				Enclosed acoustic measures & VIF             \\ \hline
				F2 mean                         & 1.86           \\
				jitter measures &2.22 \\
				shimmer measures                   &2.86       \\
				HNR                              &2.50       \\
				apq11 shimmer           &1.65       \\
				F0 mean                              &1.46       \\
				F0 SD         &1.09       \\ 
				F1 mean & 1.09\\
				F3 mean, F4 mean and VTL estimators & 1.81\\
				\hline
			\end{tabular}
			&
			\begin{tabular}{|p{0.35\columnwidth}|p{0.05\columnwidth}|}
				\hline
				\multicolumn{2}{|c|}{Male}                     \\ \hline
				Enclosed acoustic measures     &VIF       \\ \hline
				F0 SD            & 1.61                      \\
				jitter measures &1.69\\
				HNR, shimmer measures  &2.05                        \\
				apq11 shimmer             &1.42                        \\
				F3 mean, F4 mean and VTL estimators      &1.50             \\
				F2 mean                          &1.45           \\
				F1 mean&1.14\\
				F0 mean        &1.62         \\ \hline
			\end{tabular}
	\end{tabular}}
\end{table}

As the acoustic measures were merged and the VIFs were updated in each iteration loop, the 9 independent clusters for females and 8 independent clusters for males were formed at the nodes above the threshold cophenetic distance (shown as the black line in Figure \ref{fig:result_indepcluster_cluster_dendro}), corresponding to Table \ref{tab:result_indepcluster_multicol_cluster}, when the VIF values were all lower than 5. The optimal clusters and their enclosed acoustic measures included three big clusters which were formed as \{all the jitter measures\}, \{all the shimmer measures (except for apq11shimmer)\}, and \{all the VTL estimators, together with the F3 mean and F4 mean\}, as described in Section \ref{txt:result_indepcluster_cluster}. The F0 mean, F0 SD, F1 mean, F2 mean and apq11 shimmer were the five acoustic measures that were independent of other measures. The only difference for the two sexes is that HNR was grouped into the cluster of shimmer measures for males, but not for females. It is interesting that, for females, the clustering of HNR and the group of shimmer measures occurred when the number of clusters reduced to 8, which would have happened at the next clustering iteration.

As shown in Table \ref{tab:result_indepcluster_multicol_cluster}, the interpretable meanings of the independent groups of acoustic measures still hold after the 9 clusters for females and 8 clusters for males were formed for females and males respectively. The results were evaluated by reviewing the consistency or inconsistency with the published literature.
\begin{itemize}
	\item This study supports previous findings that F0 mean and VTL estimators are independent of each other \cite{feinberg2006menstrual,feinberg2005manipulations}, as they appeared in two independent clusters.
	
	\item This study is consistent with a previous study \cite{macdonald2011probing} that F0 mean, F1 mean and F2 mean are independent of each other and the other acoustic measures. The correlation between F3 mean and F4 mean was found to be strong in females (r = .75) and was moderate in males (r = .47). 
	
	\item Regarding the relationships between F3 mean and the VTL estimators,  F3 mean was highly correlated with the VTL estimators in both females and males, with r(F3 mean, $\Delta$F) = .87, r(F3 mean,avgFormant) = .87 in females, and  r(F3 mean,$\Delta$F) = .70, r(F3 mean,avgFormant) = .69 in males as shown in the original correlation matrix (\ref{app:correlation}). Consequently, F3 mean was grouped with the VTL estimators in clustering. This finding aligns with the results of \cite{monahan2010auditory} that F3 mean varies as a function of VTL across talkers, and is presented in different types of speech (e.g., whispered, nonphonated speech). Our results show that F3 mean is highly correlated with the VTL estimators which supports the conclusion drawn in a previous study \cite{claes1998novel} that F3 mean provides a good estimator of VTL in automatic speech recognition. 
	
	\item In the present study, the F4 mean was found to be more highly correlated with the VTL estimators than was the F3 mean. In females, r(F4 mean,fdisp) = .95, r(F4 mean,$\Delta$F) = .92, r(F4 mean,avgFormant) = .88, and the absolute values of correlations with the rest of the VTL estimators were all above .72. Regarding males, r(F4 mean,fdisp) = .93, r(F4 mean,$\Delta$F) = .89, r(F4 mean,avgFormant) = .81, and the absolute values of the correlations with the rest of the VTL estimators were all above .54. This evidence that F4 mean was more highly correlated with VTL estimators than was the  F3 mean, is consistent with the results of \cite{tecumseh2001descended} where VTL was strongly correlated with individual formant values, particularly for higher formants (r(F4 mean,Fitch\_f) in a range of -.49 to -.95). Therefore, the F4 mean was grouped with the VTL estimators at an earlier stage than the F3 mean, as shown in the dendrogram (Figure \ref{fig:result_indepcluster_cluster_dendro}).
	
	\item Furthermore, the F3 mean and F4 mean were found to be more highly correlated with the VTL estimators in females than in males. Specifically, as shown in the correlation matrix (see \ref{app:correlation}) in females, r(F3 mean,$\Delta$F) = .87, r(F4 mean,$\Delta$F) = .92; in males, r(F3 mean,$\Delta$F) = .70, r(F4 mean,$\Delta$F) = .89.
	
	\item To the best of our knowledge, this is the first study to statistically demonstrate that F0 SD can be regarded as independent to all other acoustic measures, with very low VIFs of 1.09 and 1.61 for females and males respectively (from Table \ref{tab:result_indepcluster_multicol_cluster}). It is noticed that F0 SD was moderately correlated with F0 mean for males (r = .41), but had a negligible correlation with F0 mean for females (r = -.15). 
	
	\item The voice perturbation measures, including HNR, jitter and shimmer, were found to be independent factors for females, but not for males, with HNR grouped with shimmer for males. However, in fact, the correlations among these three perturbation measures are still moderate rather than low for females (r(HNR, jitter) = -.63, r(HNR, shimmer) = -.66, r(jitter, shimmer) = .67, as observed in Figure \ref{fig:result_indepcluster_multicol_newr}). The reason for some correlation among these three factors was explained in study \cite{murphy1999perturbation} with reference to how the sources of periodicity perturbations can be divided into four classes: (1) pulse frequency perturbations, which is jitter, (2) pulse amplitude perturbations, which is shimmer, (3) additive noise, and (4) waveform variations, and these four classes would demonstrate correlation. Furthermore, HNR has been proposed as a measure of the amount of additive noise in the acoustic waveform. And many studies have stated HNR is not only dependent on additive noise, but also jitter and shimmer \cite{krom1993cepstrum,muta1988pitch,qi1997temporal}. These three kinds of measures may represent different voice quality, for example, shimmer and HNR can be used to classify the degree of roughness and breathiness in voice production \cite{lopes2014severity}.
\end{itemize}

\subsubsection{System Performance after Clustering}

\begin{figure}[H]
	\centering
	\includegraphics[width=\linewidth]{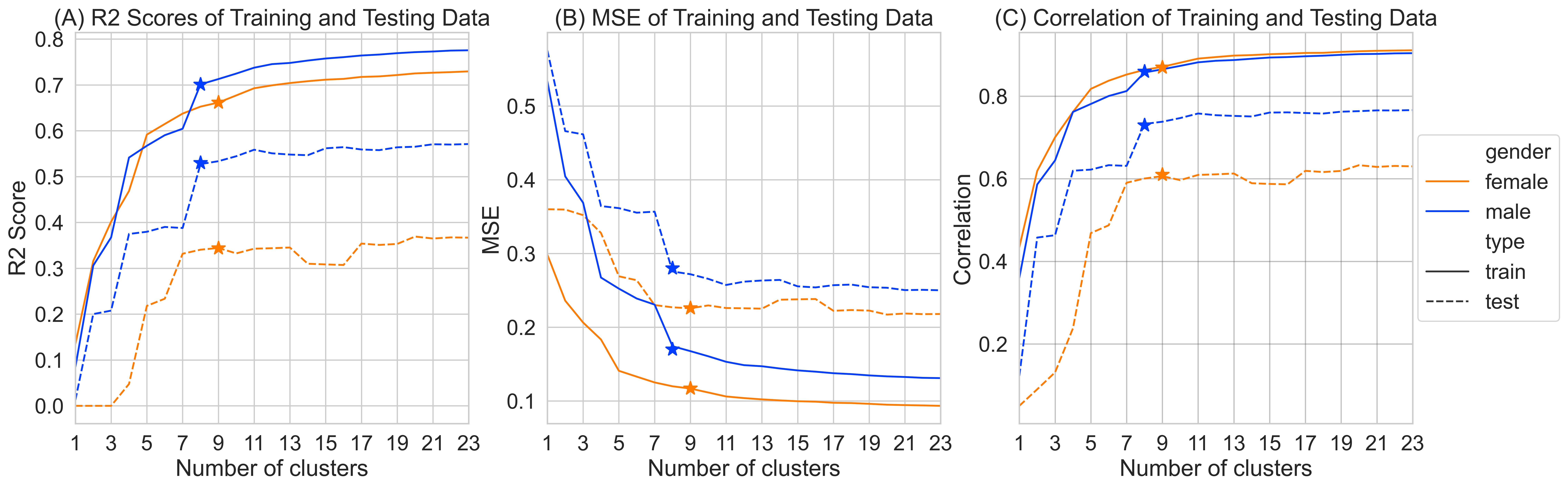}
	\caption{Relationships between model performance and number of clusters}
	\label{fig:result_indepcluster_aftercluster_ncluster}
\end{figure}

Figure \ref{fig:result_indepcluster_aftercluster_ncluster} presents the relationships between model performance ($R^2$, MSE and Correlation) and the number of clusters, with orange lines representing females and blue lines representing males. The solid lines show the performance when using training data and the dashed lines demonstrate the performance when using testing data.

Table \ref{tab:result_indepcluster_aftercluster_performance} shows the VIFs, $R^2$ scores, MSE and correlations without clustering and with clustering in males and females, using the 7-second speech data. 

\begin{table}[h!]
	\caption{VIFs, $R^2$, MSE and correlations without clustering and with clustering in males and females}
	\label{tab:result_indepcluster_aftercluster_performance}
	\centering
	\begin{tabular}{ccccccc}
		\toprule
		\textbf{Sex}&\textbf{Cluster}&\textbf{Type}&\textbf{VIFs}&\textbf{$R^2$}&\textbf{MSE}&\textbf{Correlation}\\\midrule
		\multirow{4}{*}{Female}&\multirow{2}{*}{without clustering}&training&\multirow{2}{*}{$>10^5$}&.73&.09&.91\\
		&&testing&&.37&.22&.63\\\cline{2-7}
		&\multirow{2}{*}{9 clusters}&training&\multirow{2}{*}{$<5$}&.66&.12&.87\\
		&&testing&&.34&.23&.61\\\hline
		\multirow{4}{*}{Male}&\multirow{2}{*}{without clustering}&training&\multirow{2}{*}{$>10^5$}&.78&.13&.90\\
		&&testing&&.57&.25&.77\\\cline{2-7}
		&\multirow{2}{*}{8 clusters}&training&\multirow{2}{*}{$<5$}&.70&.17&.86\\
		&&testing&&.53&.28&.73\\
		\bottomrule
	\end{tabular}
\end{table}

These results show that the best system performance was achieved when using all measures instead of using independent clusters. This is because, after clustering, data dimensionality was drastically reduced, from the original 23 dimensions to 9 dimensions for females and 8 dimensions for males respectively. However, after clustering, the redundant information carried by those highly correlated measures was eliminated, and the essential information was highly condensed within each independent group. Therefore, the $R^2$ curves in Figure \ref{fig:result_indepcluster_aftercluster_ncluster} become flat when the number of clusters increase beyond 9 for females and beyond 8 for males, and after these points the system performance differs little compared to the performance without any clustering (in Table \ref{tab:result_indepcluster_aftercluster_performance} and the end points in Figure \ref{fig:result_indepcluster_aftercluster_ncluster}).

It is interesting to point out that the results demonstrated in Figure \ref{fig:result_indepcluster_aftercluster_ncluster} are consistent with the findings reported in Section \ref{txt:result_indepcluster_multicol} that the optimal number of clusters was 9 for females and 8 for males (shown in the star points in Figure \ref{fig:result_indepcluster_aftercluster_ncluster}). We observed a dramatic drop in system performances when the number of clusters decreased from 8 to 7 for males. During this iteration, F0 mean and F0 SD were formed into one group for males. Beside this drop, it is observed in Figure \ref{fig:result_indepcluster_aftercluster_ncluster} that there was another drop from 5 clusters to 4 clusters in females, as this transition merged F0 mean and F0 SD into one cluster. It also shows a decrease from 4 clusters to 3 clusters in males, as this step merged F0 mean, F0 SD with HNR, jitter and shimmer. This behaviour suggests that masculinity/femininity rating is more dependent on the discriminative quality of the measures or the groups of measures than on the number of measures, as evident when comparing using the optimal number of highly independent groups of measures with using the 23 highly inter-dependent measures. It also implies that F0 mean, F0 SD, F2 mean and the group of F3 mean, F4 mean and the VTL estimators are the important factors in judging masculinity and femininity, as the system performance became poor when any of them was merged with other measures.

\subsection{Acoustic Characterization}
\begin{figure}[H]
	\centering
	\begin{subfigure}[b]{0.49\textwidth}
		\centering
		\includegraphics[width=1\linewidth]{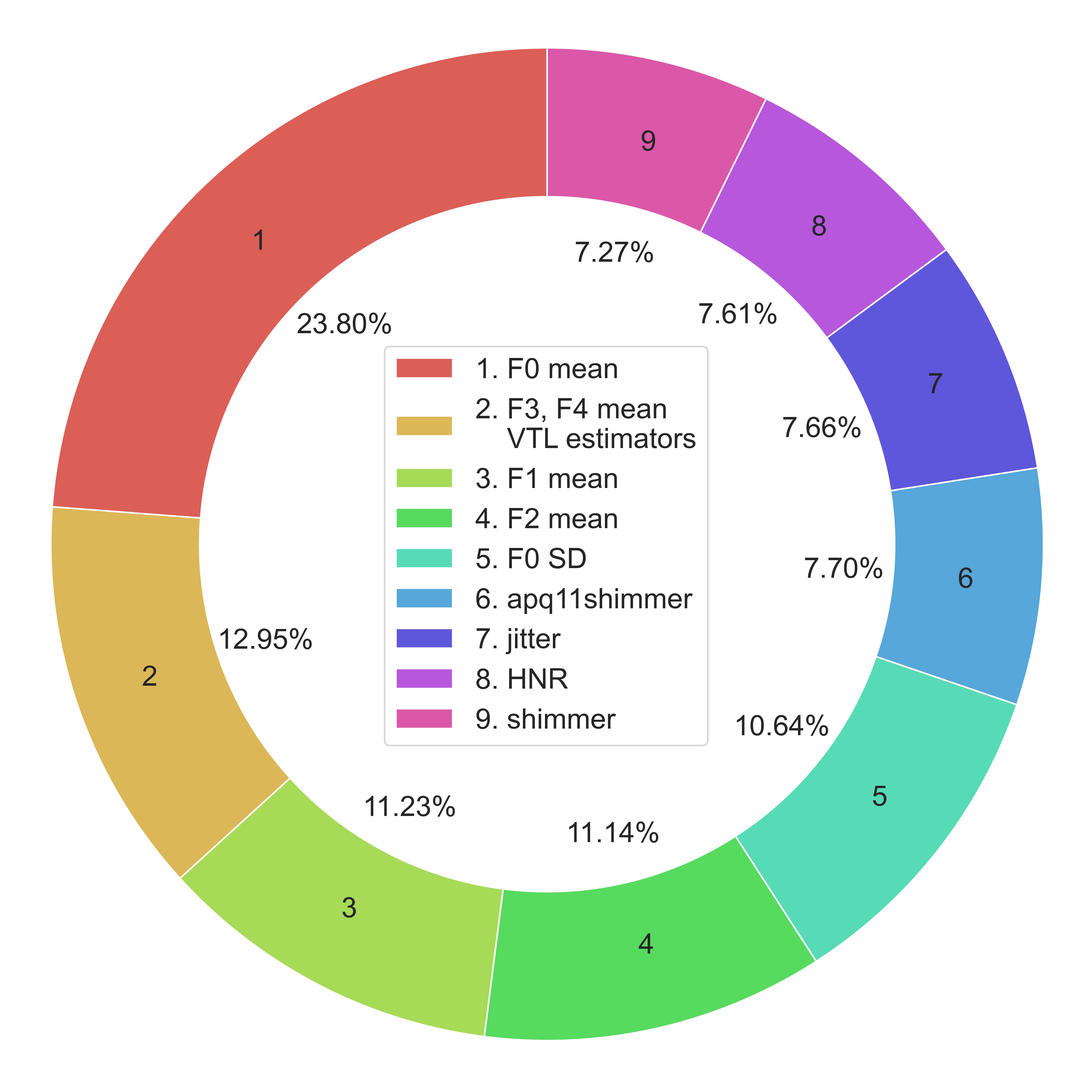}
	\end{subfigure}
	\begin{subfigure}[b]{0.49\textwidth}
		\centering
		\includegraphics[width=1\linewidth]{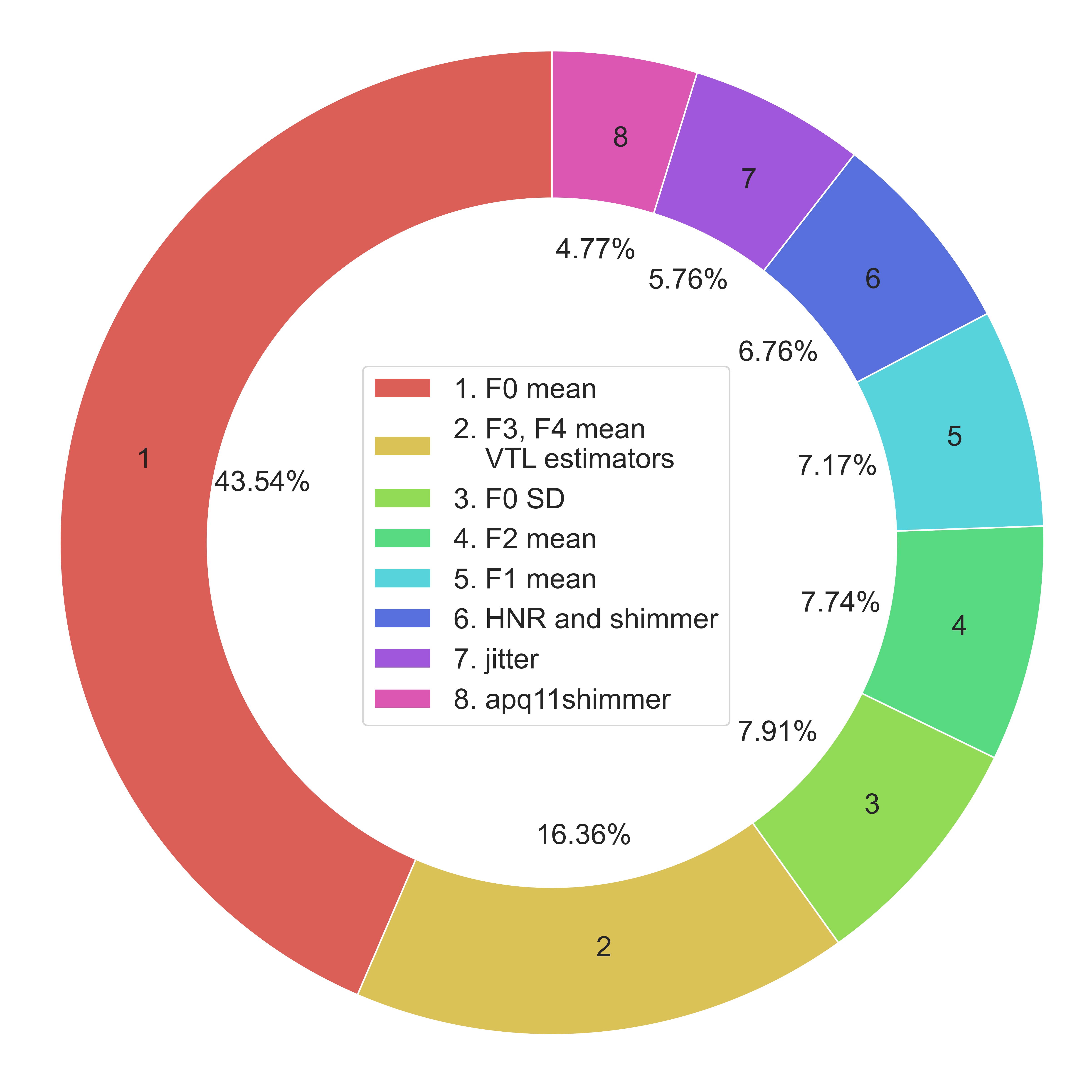}
	\end{subfigure}
	
	\caption{Cluster weights in females (left) and males (right)}
	\label{fig:result_characterization_weights}
\end{figure}

Using these independent and interpretable clusters, we next analysed their contributions in predicting the masculinity/femininity ratings. The cluster weights were obtained by applying the ERF model on the 9 clusters for female data and the 8 clusters for male data with speech duration of 7 seconds. The weights of each independent cluster are shown in Figure \ref{fig:result_characterization_weights}, where the rank of importance of each cluster is presented in a counter-clockwise way, starting from the direction of 12 o'clock, and the legend shows the rank and the enclosed acoustic measures of each cluster. 

The largest weight for F0 mean is consistent with the studies reviewed \cite{cartei2014makes,feinberg2006menstrual,little2011human,feinberg2005manipulations,pisanski2011prioritization,gelfer2000comparison,munson2007acoustic,lovato2016multi,biemans2000gender,hardy2020acoustic} that reported F0 mean to be the most critical acoustic measure in judging masculinity for males and femininity for females. However the magnitude of the contribution differs between males and females, where F0 mean plays a more important role in assessing the degree of masculinity in males (43.54\%) than in assessing the degree of femininity in females (23.8\%). 

The second important factor in masculinity/femininity rating for both males and females is the group of VTL related acoustic measures including F3 mean, F4 mean and the VTL estimators (12.95\% in females, 16.36\% in males). This result agrees with the previous literature \cite{cartei2014makes,feinberg2006menstrual,feinberg2005manipulations,pisanski2011prioritization,gelfer2000comparison,hardy2020acoustic} that these enclosed acoustic measures (F3 mean, F4 mean and VTL estimators) are correlated with perceived masculinity/femininity ratings. 

Apart from the same top two important factors in males and females, F1 mean, F2 mean and F0 SD are the next three important factors with equivalent weights ($ \approx $11\% in females, $ \approx $7.5\% in males) in the judgement of masculinity for males and femininity for females. This finding agrees with the studies \cite{pisanski2011prioritization,gelfer2000comparison,munson2007acoustic} that F1 mean and F2 mean are significant and independent predictors of perceived masculinity in males and perceived femininity in females. However, this finding contrasts with the finding of study \cite{gelfer2000comparison} that F2 mean was statistically more significant in accounting for perceived masculinity/femininity ratings than either F1 mean or F3 mean. Actually, it was found in the present study that the weights of F2 mean and F1 mean were equivalent no matter in predicting femininity for females (11.14\% v.s. 11.23\%) or in predicting masculinity for males (7.74\% v.s. 7.17\%). Further, F3 mean, together with the other highly correlated acoustic measures (F4 mean and the VTL estimators), weight more than either F2 mean or F1 mean in the masculinity/femininity rating prediction, especially in the judgement of femininity. Surprisingly, the present study is the first to find the importance of F0 SD in predicting masculinity/femininity ratings. The F0 SD, as one of the independent factors, is as critical as F1 mean and F2 mean in accounting for variance in masculinity/femininity ratings. F0 SD has not been paid much attention in the previous research regarding the classification of gender for speech or the assessments of masculinity/femininity.

Finally, vocal perturbation related acoustic measures, which include HNR, shimmer measures and jitter measures, are the least important factors in predicting masculinity/femininity ratings. This finding aligns with studies \cite{owen2010role,king2012voice} in suggesting that perturbation measures do not affect perceived masculinity or femininity. According to studies \cite{teixeira2014jitter,biemans2000gender}, perturbation measures are able to differentiate between males and females because the female voice has significantly greater jitter and less shimmer than male voice. This between-class difference was validated by Andrews and Schmidt \cite{andrews1997gender} who reported that voice samples produced by females were perceived to be more breathy and less hoarse than those produced by males, and breathiness has been associated with increased jitter \cite{king2012voice} and shimmer values have been suspected to be a measure of vocal hoarseness. However, it seems that jitter and shimmer may only be regarded as discriminators between the two classes of males and female, but not salient factors that influence the perceived masculinity for males or perceived femininity for females. 

\subsection{Limitations}
In the present study, we chose to use a particular sentence of read text because of the following reasons. Firstly, this was expected to minimize sources of variation in speech that would come from using spontaneous speech. For instance, the semantic content of the speech would vary across speakers for spontaneous speech and this content may influence perceived masculinity/femininity. Secondly, the stimulus type used in our study focused on the mechanics of voice production and control of each speaker under a general scenario, rather than concentrating on the paralinguistic characteristics of each speaker under various scenarios. For instance, numerous studies have shown that F0 could vary significantly when the speaker expressed different emotions in their speech \cite{razak2003emotion,breiman2001random}. Lastly, read speech was chosen instead of spontaneous speech based on studies \cite{furui2005recognition,nakamura2008differences,pfitzinger1996syllable} that showed spontaneous speech reduces recognition accuracy for phonemes, consonants and syllables, which implies a corresponding reduction in the discriminating acoustic information. However, considering human raters may use paralinguistic features, such as variation of tone and emotion status, to assess the speakers’ masculinity/femininity under different scenarios, spontaneous speech or conversational speech could be used to analyze the paralinguistic features affecting perceived masculinity/femininity. 

Additionally, all participants of our study were Caucasian living in Australia who were self-reported males or females defined with reference to their sex as assigned at birth, therefore our model would be expected to perform at its best for an unknown speaker with the same background. Future studies could also consider some important factors that could influence the judgement of perceived masculinity/femininity, such as the ethnicity and age of both speakers and listeners. We acknowledge that our model considered only the binary biological sexes. As perceived vocal masculinity/femininity could also be influenced by individual gender identities, further work needs to be conducted to develop appropriate models to account for participants’ gender identities. 

\section{Conclusions}
This study investigated a novel model framework with the following objectives: (1) rating the speakers' acoustic-related masculinity and femininity on a set of acoustic measures using a machine learning model, (2) building independent groups of acoustic measures with interpretable physical meanings to eliminate multicollinearity, and (3) characterization of the salient acoustic measures associating with the perceived masculinity/femininity ratings. The model provided promising results that the machine ratings of masculinity/femininity are strongly correlated with the perceived masculinity/femininity ratings when using the data with the optimal speech duration of 7 seconds. Moreover, the model built 9 and 8 independent meaningful groups of acoustic measures for females and males respectively which showed no worse performance in predicting the perceived masculinity/femininity ratings than using the 23 acoustic measures before clustering. Based on the optimal speech duration and the independent meaningful groups of acoustic measures, the model provided the importance of the groups in predicting the perceived masculinity and femininity. The results revealed that F0 mean and the group of F3 mean, F4 mean and VTL estimators are the top two characteristics that affect the judgement of speakers' masculinity and femininity, with F0 mean being the most significant factor in assessing the masculinity in males. The F1 mean, F2 mean and F0 standard deviation share similar importance, and the voice perturbation measures, including HNR, jitter and shimmer are the least important. 




\bibliographystyle{elsarticle-num-names}
\bibliography{SPECOM}







\appendix
\section{Pseudocode of Machine Rating of Masculinity/Femininity }
\label{app:pseudogenderrating}
\resizebox{1\columnwidth}{!}{
	\begin{algorithm}[H]
		\KwData{23 acoustic measures extracted from 1/2/5/7/10 seconds segments and 2nd sentence segments for both males and females, together with perceived masculinity/femininity ratings, 12 sets of data in total}
		\KwIn{One set of Data}
		\KwOut{$R^2_{train}$, $R^2_{test}$, ${MSE}_{train}$, ${MSE}_{test}$, $r_{train}$ and $r_{test}$}
		\tcc{Data size = m samples $\times$ 23 acoustic measures, m = number\ of\ participants x total speech duration per participants/speech duration}
		\SetKw{paramgrid}{param\_grid}
		\SetKw{Rtrain}{$R^2_{train}$}
		\SetKw{Rtest}{$R^2_{test}$}
		\SetKw{MSEtrain}{${MSE}_{train}$}
		\SetKw{MSEtest}{${MSE}_{test}$}
		\SetKw{rtrain}{$r_{train}$}
		\SetKw{rtest}{$r_{test}$}
		\SetKw{Rtrainkf}{$R^2_{train}kf$}
		\SetKw{Rtestkf}{$R^2_{test}kf$}
		\SetKw{MSEtrainkf}{${MSE}_{train}kf$}
		\SetKw{MSEtestkf}{${MSE}_{test}kf$}
		\SetKw{rtrainkf}{$r_{train}kf$}
		\SetKw{rtestkf}{$r_{test}kf$}
		\SetKw{in}{in}
		\SetKw{X}{X}
		\SetKw{Y}{Y}
		\Begin{
			\X,\Y = StandardScalar(input);\\
			\tcc{Standardize data by removing the mean and scaling to unit variance.}
			kf = KFold (n\_splits = 4, shuffle = True, on = IDs\_speaker);\\
			\tcc{4 Folds cross validation, splitting dataset into 4 consecutive folds with shuffling, the training data and testing data were split based on individual speaker.}
			\paramgrid = \{'max\_depth','min\_samples\_leaf','min\_samples\_split','max\_leaf\_nodes'\};\\
			ERF = ExtraTreesRegressor(n\_estimator = 1000)
			\tcc*{initiate an ERF model}
			clf = GridSearchCV(ERF,\paramgrid,cv=10);\\
			\tcc{exhaustive search over specified parameter values, cross-validation splitting strategy of 10 folds, max\_depth = 10, max\_leaf\_nodes = 300, min\_samples\_leaf = 2}
			clf.fit(\X);\\ 
			\tcc{train the ERF model on the entire input dataset to get the best fit hyper-parameters}
			best\_param = clf.best\_params\_;\\
			ERF\_best = ExtraTreesRegressor(n\_estimator = 1000, param = best\_param);\\
			\For{train, test \in kf.split(\X,\Y)}{
				ERF\_best.fit(train);\\ 
				\tcc{apply the best fit hyper-parameters on the training dataset}
				train\_prediction = ERF\_best(train);\\
				test\_prediction = ERF\_best(test);\\
				\Rtrainkf.append(R2(train,train\_prediction));\\
				\Rtestkf.append(R2(test,test\_prediction));\\
				\MSEtrainkf.append(mean\_squared\_error(train,train\_prediction));\\
				\MSEtestkf.append(mean\_squared\_error(test,test\_prediction));\\
				\rtrainkf.append(pearsonr(train,train\_prediction));\\
				\rtestkf.append(pearsonr(test,test\_prediction));\\
				\tcc{intermediate output, each one contains 4 values from 4 folds cross validation}
			}
			\Rtrain,\Rtest,\MSEtrain,\MSEtest,\rtrain,\rtest = mean(\Rtrainkf,\Rtestkf,\MSEtrainkf,\MSEtestkf,\rtrainkf,\rtestkf)
		}	
		\caption{Pseudo code of machine rating of masculinity/femininity}
		\label{pseudocode1}	
	\end{algorithm}
}

\section{Pseudocode of Hierarchical Clustering Algorithm}
	\label{app:pseudocluster}
	\resizebox{1\columnwidth}{!}{
	\begin{algorithm}[H]
	\SetKw{in}{in}
	
	\KwData{Correlation Matrix of 23 Acoustic Measures, size of 23 $\times$ 23, with labels as $ m_1, m_2, \dots, m_{23} $}
	\KwOut{Stepwise dendrogram; breakdown data structure of dendrogram - L, an (N-1)$\times$3 - matrix}
	\SetKw{Def}{Def}
	\Begin{
	\Def dist($ c_1,c_2 $) \tcc*{A distance function Eq \ref{eq:method_buildindepclusters_cluster_duv}}
	\For{i =1 to 23}{$ c_i = {m_i} $}
	C = {$ c_1 $, \dots, $ c_{23} $}\\
	\While{C.size $ > $ 1}{
	($ c_{min1},c_{min2} $) = minimum dist($ c_i,c_j $) for all $ c_i $, $ c_j $ \in C\\
	remove $ c_{min1}$ and $c_{min2} $ from C\\
	add $ c_{new} = \{c_{min1},c_{min2}\} $ to C\\
	L.append($ c_{min1},c_{min2},dist(c_{min1},c_{min2}) $)
	}
	return L
	}	
	\caption{Hierarchical clustering algorithm}
	\label{alg:clustering}	
	\end{algorithm}
	}

\section{Pseudo Code of Building Independent Clusters and Characterization}
\label{app:character}
\begin{center}
	
	\resizebox{1\columnwidth}{!}{
		\centering
		\begin{algorithm}[H]
			
			\footnotesize
			\KwData{23 acoustic measures extracted from the segments with the optimal speech duration for both males and females, together with perceived gender ratings}
			\KwIn{One set of Data}
			\KwOut{VIFs, $R^2_{train}$, $R^2_{test}$, ${MSE}_{train}$, ${MSE}_{test}$, $r_{train}$ and $r_{test}$, cluster\_importance}
			\tcc{Data size = m samples $\times$ 23 acoustic measures, m = number\ of\ participants x total speech duration per participants/ the optimal speech duration}
			\SetKw{ncluster}{n\_cluster}
			\SetKw{clusters}{clusters}
			\SetKw{dendrogram}{dendrogram}
			\SetKw{clusteri}{cluster\_i}
			\SetKw{measurei}{measure\_i}
			\SetKw{Xmeasurei}{X[measure\_i]}
			\SetKw{newXclusteri}{new\_X[cluster\_i]}
			\SetKw{newX}{new\_X}
			\SetKw{rmatrix}{r\_matrix}
			\SetKw{VIFs}{VIFs}
			\SetKw{in}{in}
			\SetKw{newrmatrix}{new\_r\_matrix}
			\SetKw{paramgrid}{param\_grid}
			\SetKw{Rtrain}{$R^2_{train}$}
			\SetKw{Rtest}{$R^2_{test}$}
			\SetKw{MSEtrain}{${MSE}_{train}$}
			\SetKw{MSEtest}{${MSE}_{test}$}
			\SetKw{rtrain}{$r_{train}$}
			\SetKw{rtest}{$r_{test}$}
			\SetKw{Rtrainkf}{$R^2_{train}kf$}
			\SetKw{Rtestkf}{$R^2_{test}kf$}
			\SetKw{MSEtrainkf}{${MSE}_{train}kf$}
			\SetKw{MSEtestkf}{${MSE}_{test}kf$}
			\SetKw{rtrainkf}{$r_{train}kf$}
			\SetKw{rtestkf}{$r_{test}kf$}
			\SetKw{X}{X}
			\SetKw{Y}{Y}
			\SetKw{clusterimportance}{cluster\_importance}
			\Begin{
				\X,\Y = StandardScalar(input);\\
				\tcc{Standardize data by removing the mean and scaling to unit variance.}
				\rmatrix = abs(\X.corr());\\
				\tcc{absolute values of pairwise correlations of clusters, initial size of 23x23}
				\dendrogram = Hierarchical Clustering Algorithm (data = \rmatrix, method='average');\\
				\For{\ncluster $\leftarrow$ 23 \KwTo 2 \tcc*{number of clusters, 2$\leq$measure\_i$\leq$23}}{
					\VIFs = vif(\rmatrix) \tcc*{a list of VIF values, the list length = \ncluster}
					\clusters = \dendrogram(n=\ncluster);\\
					\tcc{generate a list of clusters, the length of each item in it is \ncluster}
					\For{cluster \in \clusters}{
						\If{length(cluster)$>$1}{\newXclusteri = PCA(n\_component=1)(sample[\measurei for \measurei in cluster])}
						\tcc{\clusteri - an element of clusters, one cluster may contain one or multiple acoustic measures, the combination of all clusters covers 23 measures with no repetition\\ \measurei - the index number of i\_th acoustic measure, 0$\leq$measure\_i$<$23\\ \newXclusteri - the values of the i\_th cluster in samples, size of m samples$\times$1}
						\Else{\newXclusteri = \Xmeasurei}
						\tcc{\Xmeasurei - the values of the i\_th acoustic measure in samples, size of m samples$\times$1; \newX size = m samples$\times$\ncluster}
					}
					\rmatrix = abs(\newX.corr()) \tcc*{update \rmatrix}
					kf = KFold (n\_splits = 4, shuffle = True, on = IDs\_speaker);\\
					\tcc{4 Folds cross validation, splitting dataset into 4 consecutive folds with shuffling, the training data and testing data were split based on individual speaker.}
					\paramgrid = \{'max\_depth','min\_samples\_leaf','min\_samples\_split','max\_leaf\_nodes'\};\\
					ERF = ExtraTreesRegressor(n\_estimator = 1000)
					\tcc*{initiate an ERF model}
					clf = GridSearchCV(ERF,\paramgrid,cv=10);\\
					\tcc{exhaustive search over specified paramter values, cross-validation splitting strategy of 10 folds}
					clf.fit(\newX);\\ 
					\tcc{train the ERF model on the entire new samples to get the best fit hyper-parameters}
					best\_param = clf.best\_params\_;\\
					ERF\_best = ExtraTreesRegressor(n\_estimator = 1000, param = best\_param);\\
					\For{train, test \in kf.split(\newX,\Y)}{
						ERF\_best.fit(train);\\ 
						\tcc{apply the best fit hyper-parameters on the training dataset}
						train\_prediction = ERF\_best(train);\\
						test\_prediction = ERF\_best(test);\\
						\Rtrainkf.append(R2(train,train\_prediction));\\
						\Rtestkf.append(R2(test,test\_prediction));\\
						\MSEtrainkf.append(mean\_squared\_error(train,train\_prediction));\\
						\MSEtestkf.append(mean\_squared\_error(test,test\_prediction));\\
						\rtrainkf.append(pearsonr(train,train\_prediction));\\
						\rtestkf.append(pearsonr(test,test\_prediction));\\
						\tcc{intermediate output, each one contains 4 values from 4 folds cross validation}
					}
					\Rtrain,\Rtest,\MSEtrain,\MSEtest,\rtrain,\rtest = mean(\Rtrainkf,\Rtestkf,\MSEtrainkf,\MSEtestkf,\rtrainkf,\rtestkf)
				}
				Determine on the optimal \ncluster\;
				\clusterimportance = ERF\_best(\ncluster).feature\_importance\_
			}	
			\caption{Pseudo code for building independent clusters and characterization}
			\label{pseudocode}	
		\end{algorithm}
	}
\end{center}

\section{Correlation Matrix of Unclusterd Acoustic Measures}
\label{app:correlation}
\begin{figure}[H]
	\centering
	\begin{subfigure}[b]{0.8\textwidth}
		\centering
		\includegraphics[width=1\linewidth]{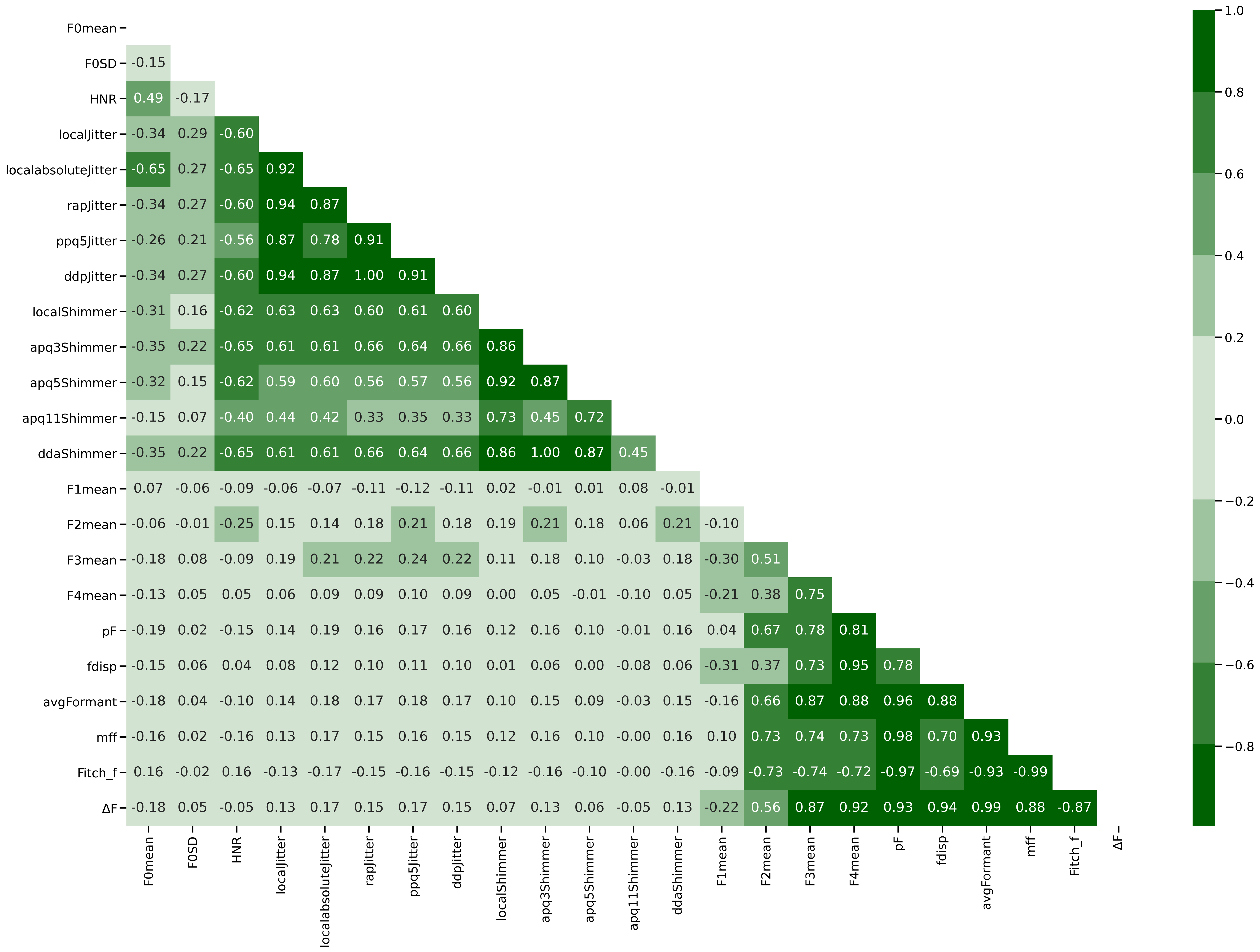}
	\end{subfigure}
	\qquad
	\begin{subfigure}[b]{0.8\textwidth}
		\centering
		\includegraphics[width=1\linewidth]{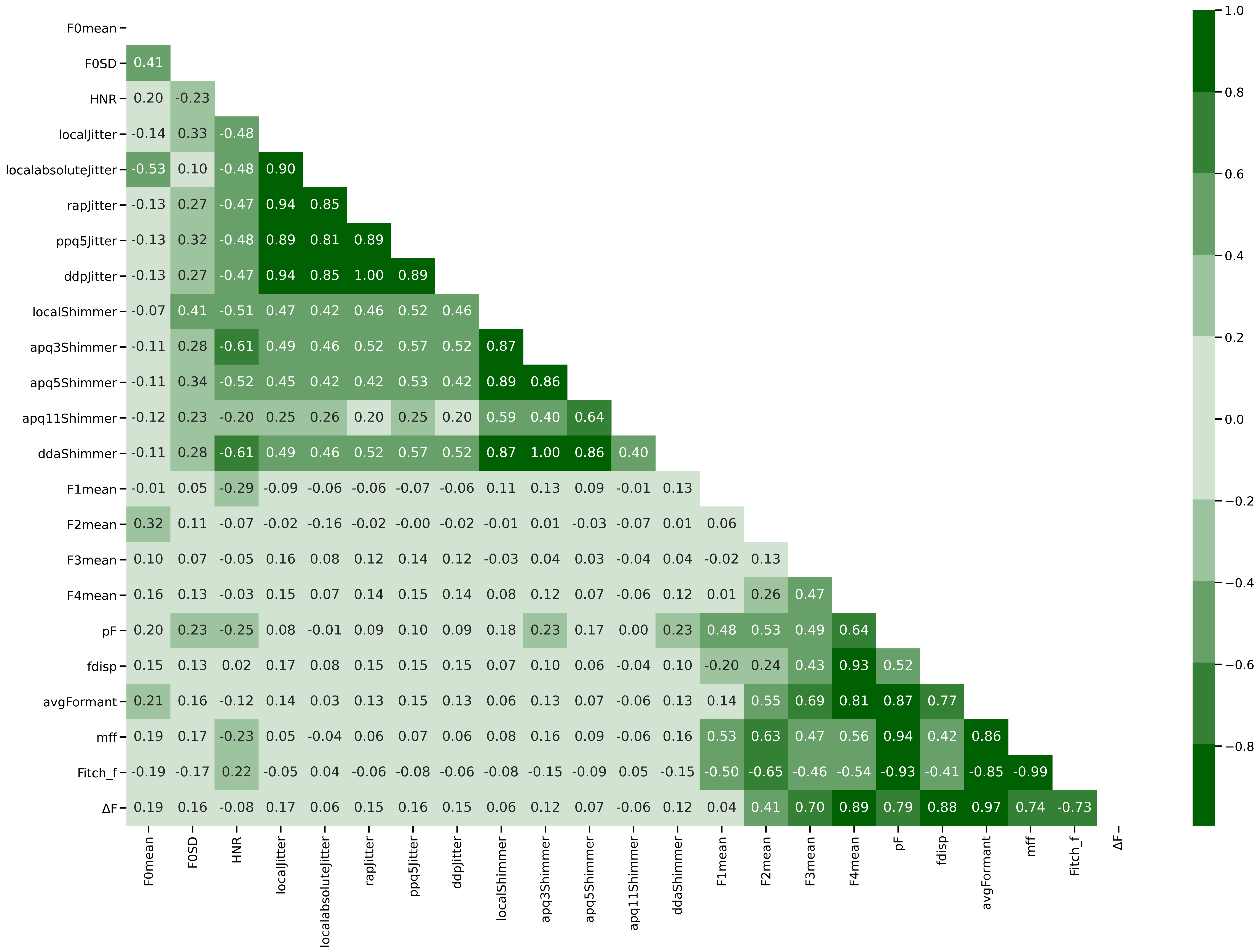}
	\end{subfigure}
	
	\caption{Correlation matrix of unclustered acoustic Measures in females (top) and males (bottom)}
	\label{fig:bigcorm}
\end{figure}


\end{document}